\documentclass[acmsmall]{acmart}

\AtBeginDocument{%
  }

\setcopyright{acmlicensed}
\acmDOI{10.1145/3643752}
\acmYear{2024}
\copyrightyear{2024}
\acmSubmissionID{fse24main-p390-p}
\acmJournal{PACMSE}
\acmVolume{1}
\acmNumber{FSE}
\acmMonth{7}

\usepackage{tabularx}
\usepackage{graphicx}
\usepackage{subcaption}
\usepackage{multirow}
\usepackage[export]{adjustbox}
\usepackage{xcolor}
\usepackage{colortbl}
\usepackage[export]{adjustbox}
\usepackage{xspace}

\begin{document}

\newcommand{\toolname}{{\sc CodeArt}\xspace}

\newcommand{\revise}[1]{{\color{black}{#1}}}
\newcommand{\rebuttal}[1]{{\color{black}{#1}}}
\newcommand{\rebuttalshort}[1]{{\color{black}{#1}}}
\definecolor{Gray}{gray}{0.935}
\newcolumntype{g}{>{\columncolor{Gray}}c}
\newcolumntype{G}{>{\columncolor{Gray}}r}
\title{CodeArt: Better Code Models by Attention Regularization When Symbols Are Lacking}

\author{Zian Su}
\orcid{0009-0001-8283-2825}
\affiliation{%
  \institution{Purdue University}
  \city{West Lafayette}
  \country{USA}
}
\email{su284@purdue.edu}

\author{Xiangzhe Xu}
\orcid{0000-0001-6619-781X}
\affiliation{%
  \institution{Purdue University}
  \city{West Lafayette}
  \country{USA}
}
\email{xu1415@purdue.edu}

\author{Ziyang Huang}
\orcid{0009-0009-2764-6091}
\affiliation{%
  \institution{Purdue University}
  \city{West Lafayette}
  \country{USA}
}
\email{huan1562@purdue.edu}

\author{Zhuo Zhang}
\orcid{0000-0002-6515-0021}
\affiliation{%
  \institution{Purdue University}
  \city{West Lafayette}
  \country{USA}
}
\email{zhan3299@purdue.edu}

\author{Yapeng Ye}
\orcid{0000-0001-7232-0650}
\affiliation{%
  \institution{Purdue University}
  \city{West Lafayette}
  \country{USA}
}
\email{ye203@purdue.edu}

\author{Jianjun Huang}
\orcid{0000-0003-4403-0060}
\authornote{Corresponding author.}
\affiliation{%
  \institution{Renmin University of China}
  \city{Beijing}
  \country{China}
}
\email{hjj@ruc.edu.cn}

\author{Xiangyu Zhang}
\orcid{0000-0002-9544-2500}
\affiliation{%
  \institution{Purdue University}
  \city{West Lafayette}
  \country{USA}
}
\email{xyzhang@cs.purdue.edu}

\begin{abstract}
Transformer based code models have impressive performance in many software engineering tasks. However, their effectiveness degrades when symbols are missing or not informative. The reason is that the model may not learn to pay attention to the right correlations/contexts without the help of symbols. We propose a new method to pre-train general code models when symbols are lacking. 
We observe that in such cases, programs degenerate to something written in a very primitive language. We hence propose to use program analysis to extract contexts a priori (instead of relying on symbols and masked language modeling as in vanilla models). We then leverage a novel attention masking method to only allow the model attending to these contexts, e.g., bi-directional program dependence transitive closures and token co-occurrences.
In the meantime, the inherent self-attention mechanism is utilized to learn which of 
the allowed attentions are more important compared to others.
To realize the idea, we enhance the vanilla tokenization and model architecture of a BERT model, construct and utilize attention masks,
and introduce a new pre-training algorithm.
We pre-train this BERT-like model from scratch, using
a dataset of 26 million stripped binary functions with explicit program dependence information extracted by our tool. We apply the model in three downstream tasks: binary similarity, type inference, and malware family classification. Our pre-trained model can improve the SOTAs in these tasks from 53\% to 64\%, 49\% to 60\%, and 74\% to 94\%, respectively. 
It also substantially outperforms other general pre-training techniques of code understanding models.
\end{abstract}



\maketitle

\section{Introduction}
Transformer models
~\cite{devlin2018bert,raffel2020exploring,brown2020language,achiam2023gpt}
have substantially advanced the state-of-the-art of Natural Language Processing (NLP) applications. They are also the technique behind Large Language Models (LLMs) which have demonstrated unprecedented generalization and reasoning ability. These models feature the {\em self-attention mechanism}~\cite{vaswani2017attention} that allows them to learn correlations/contexts distant in the input space and self-supervised pre-training methods, such as {\em Masked Language Modeling} (MLM) that can produce high quality embeddings by masking parts of inputs and forcing the models to predict the masked parts from their contexts.
Although Transformer models were initially introduced to improve NLP applications, recent research has shown that they can be used in many software engineering tasks, such as automated program repair~\cite{peng2023domain,wu2023effective,cao2023study,fan2023automated}, software testing~\cite{fang2023isyn}, vulnerability detection~\cite{steenhoek2024dataflow,napier2023empirical} and so on~\cite{lin2023cct5,xu2022systematic,fang2023representthemall,guo2023study}, outperforming traditional methods. An underlying reason of their impressive performance is that software has rich natural language artifacts (also called {\em symbols} in this paper for terminology simplicity), such as comments, documents, variable and function names. These artifacts make programs ``understandable'' to Transformer models, just like to human developers. For example, the correlation between a statement that defines a variable and another statement that uses the variable can be naturally captured by the attention mechanism due to the common variable name, just like how developers infer dataflow by variable names. 

However, when symbols are not available, such as in stripped binary executables, or not informative, such as in obfuscated software, programs become extremely difficult for models to understand, just like they are hard for developers. 
In particular, they are in a very primitive language in which tokens no longer have rich semantics. For example, in x86 executables, variables are denoted by registers and memory locations dereferenced through registers or constants; a same register may be allocated to multiple variables. As such, the definition of a register and the use of the register may not suggest dependence; neighboring tokens may belong to completely independent contexts/computations. Such context interleavings are difficult to unfold without the help of symbols.
As shown by our experiments, code models based on vanilla Transformer architectures and using the vanilla MLM pre-training have degraded performance when symbols are lacking.

To mitigate the problem, researchers have proposed various methods. Trex~\cite{pei2020trex}
used micro-executions to acquire input and output operand values of an x86 instruction and then leveraged such values to train a Transformer model to 
precisely represent instruction level semantics. JTrans~\cite{wang2022jtrans} used jump target prediction together with MLM to pre-train a Transformer model and used contrastive learning to force the model to learn embeddings that can distinguish similar and dissimilar binary functions.
DiEmph~\cite{xu2023diemph} further improved JTrans by removing biases (e.g., undesirable code pattern distributions) introduced by compilers.
GraphCodeBERT~\cite{guo2020graphcodebert} 
aimed to enhance Transformer code model pre-training leveraging dataflow information. 
It expanded the raw input sequence with additional tokens denoting variables and introduced extra training losses to force the model to learn data-flow between these tokens. 
While these proposals demonstrated great improvements over vanilla models, 
most of them focused on single downstream tasks, instead of general pre-trained models. Some (e.g., GraphCodeBERT) required symbols.
Note that there were also a body of works that treated programs as graphs (e.g., control flow graphs and dependence graphs) and leveraged {\em Graph Neural Networks} (GNNs) for software engineering tasks~\cite{gao2024sigmadiff,kim2022improving,yu2022deepdi,dong2022fira,sun2020treegen}. %
However, their performance also degrades when symbols are lacking, e.g., due to their need of labeled data in supervised learning and difficulties in capturing correlations that are multiple edges away in graphs
~\cite{wu2021representing}. More can be found in Section~\ref{sec:motivation}.%

In this paper, we aim to develop a new technique for pre-trained general code models that targets programs without meaningful symbols, binary executables in particular. It does not fine-tune an existing pre-trained model. Instead, it pre-trains a model from scratch using a new method that regularizes attention. It produces high quality embeddings that encode program dependences and disentangle interleaving contexts and hence enables better performance in downstream tasks. Its intuition is the following. Without meaningful symbols, a programming language degenerates to a very primitive one like an arithmetic language~\cite{pierce2002types}, in which the meaning of a variable/statement can only be derived from the direct and transitive computations that produce and use the value of the variable/statement. For example, in a function where all statements in the function body cohesively compute a final return value, the embedding of the return value should reflect the computation of the whole function; a function computing multiple orthogonal output values shall have embeddings reflecting such orthogonality in spite of interleavings of the sub-computations. The language is  dissimilar to a natural language,
and humans rarely speak in such a primitive fashion.
As such, the vanilla MLM method can hardly help the model produce the right attention during pre-training without the help of symbols because MLM mainly leverages the correlations between individual tokens and their left and right contexts
(shown in Section~\ref{sec:motivation}).

To address the challenge, we propose to use program analysis to derive possible dependences between instructions and then construct {\em attention masks} from such dependences. The masks enable self-attention between instructions that have dependences and preclude attention among those that are independent. This aligns well with the aforementioned primitive language.
The pre-training then helps the model determine which dependences %
are more important than others and hence deserve more attention. 
Besides the dependence masks, additional masks are created to explicitly regularize self-attention 
to correlations other than program dependences, such as token co-occurrences.
We also enhance the MLM pre-training by masking part of program dependences and introducing spurious dependences, and then forcing the model to correctly predict/classify such dependences. During inference, \toolname takes the subject binary and generates the corresponding attention masks and feeds both the input tokens and the masks to produce output embeddings. Note that the dependence analysis and mask construction are deterministic and transparent to users. 
Our contributions are summarized in the following.

\begin{itemize}
\item We propose a new method to pre-train Transformer code models when symbols are lacking. Inspired by existing works that utilize program analysis to enhance Deep Learning models~\cite{li2022soft,gao2023two,bui2021treecaps,shi2022better,ding2023traced,shi2023towards,chen2022transferability,saberi2023model,ma2022graphcode2vec,bui2021infercode,cao2022galois,ding2023towards,chakraborty2022natgen,wang2022bridging,ding2022can,sattler2023seal},
our method analyzes program dependences and use them to help self-attention. Different from many existing techniques that focus on improving performance of individual downstream tasks, our pre-trained models are general, serving a large number of applications, and use masks to regulate attention.

\item We address a number of technical challenges, including enhancements of tokenization, model architecture, a new pre-training method that masks dependences, transforming transitive dependences to connectivity relations to avoid undesirable decay,
and new training objectives. 
Built on top of the BinaryCorps dataset~\cite{wang2022jtrans}, we construct a large-scale training dataset with 26 million stripped binary functions containing explicit dependence information. 

\item We develop a prototype \toolname ({\it Better CODE models by Attention regulaRizaTion when symbols are lacking})
and use it to pre-train a BERT-like general model from scratch. The pre-training converges in four days with 
an 8×A100 GPU cluster.
To demonstrate the generalization of the pre-trained model, we use it in three downstream tasks: binary similarity analysis, malware classification, and binary type inference. We have improved the SOTAs of these tasks from 53\% to 64\% (Recall@1 with a pool size of 500), from 49\% to 60\% (LRAP), and from 74\% to 94\% (F-1 score averaged over different optimizations), respectively. We empirically compare with other general code pre-training approaches like GraphCodeBERT on binary code
and show that our model is much more effective. We also conduct an ablation study to justify our design choices. ~\footnote{Code and data are available at \href{https://github.com/ziansu/codeart}{https://github.com/ziansu/codeart}.}

\end{itemize}

\section{Motivation}

\label{sec:motivation}

We use an example to discuss the limitations of existing techniques and illustrate our technique.
For better readability, we present the example mostly in its source code form. Our technique works on stripped binaries without any symbol information. 

\begin{figure}
    \centering
    \includegraphics[width=\linewidth]{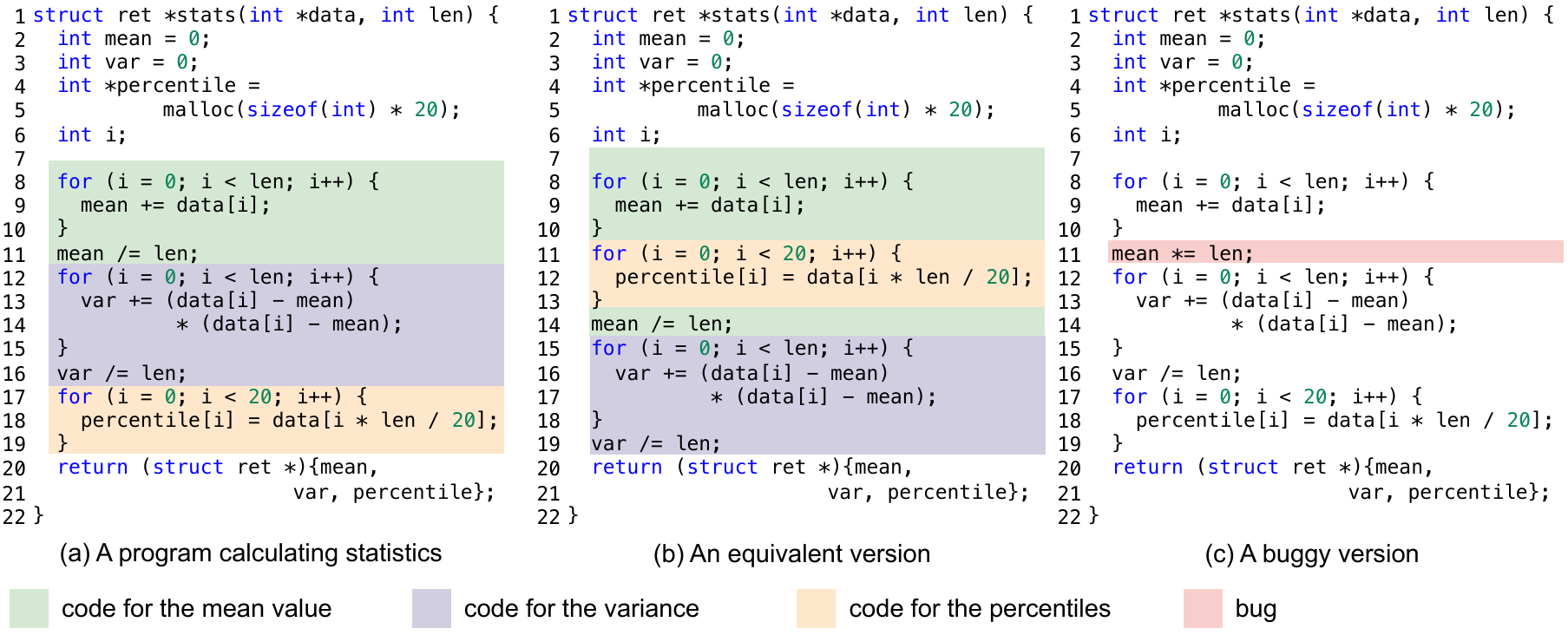}
    \caption{Code examples calculating statistics: (a) and (b) are equivalent but have different statement orders; (c) is buggy at line 11 with the wrong operation.}
    \label{fig:moti-ex}
\end{figure}

\smallskip
\noindent
{\bf Example.} Fig.~\ref{fig:moti-ex}(a) shows a code snippet computing mean, variance, and percentile for an array of sorted data. The code blocks in different colors denote the three sub-computations. Fig.~\ref{fig:moti-ex}(b) shows a version equivalent to (a) with statements reordered. Fig.~\ref{fig:moti-ex}(c) is highly similar to (a) except that its line 11 has a buggy operation. We mix (b) and (c) with a set of 5 random functions to form a pool of candidate functions and use (a) to query the most similar function from the pool. We first use a similarity analysis built on the CodeT5 model using source code. 
The analysis easily identifies that (a) is similar to (b) and not to (c). However, 
when we use a binary similarity analysis JTrans that operates on stripped binaries (without symbols), JTrans mistakenly considers (a) is similar to (c) instead of (b). 
In contrast, a similarity analysis built on our model reports the correct similarity result even without symbols. In the following, we explain these different results. 

\smallskip
\noindent
{\bf Transformer Models.}
Recent research has shown that Transformer based code models, including Large Language Models (LLMs), deliver superior performance in many software engineering tasks~\cite{feng2020codebert,lu2021codexglue,wang-etal-2021-codet5,roziere2023code}. %
Transformer was initially introduced for NLP applications~\cite{vaswani2017attention,devlin2018bert,raffel2020exploring,radford2018improving}, and 
the rich natural language artifacts in software make it suitable for these tasks.
However, in tasks where symbols are precluded (e.g., binary analysis) or not informative (e.g., in obfuscated software), code becomes dissimilar to natural language products, causing the model to have sub-optimal attention and hence performance degradation.

\begin{figure}
    \centering
    \includegraphics[width=.94\linewidth]{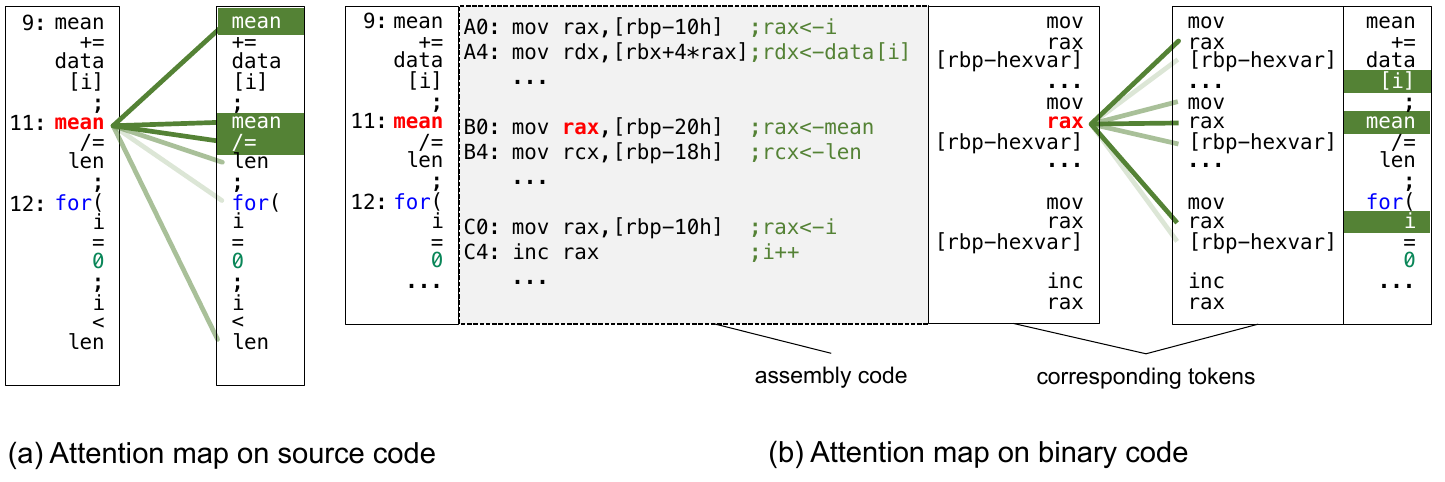}
    \caption{Attention maps of variable \texttt{mean} at line 11 of Fig.~\ref{fig:moti-ex}(a) (in red), by a source code model (a) and by a binary code model (b).
    In (a), we show attention between source code tokens, corresponding to lines~9--11 in Fig.~\ref{fig:moti-ex}(a).
    In (b), we show attention between binary code tokens. For readability, we also include the corresponding source code and part of the assembly. 
    The variable \textit{mean} is stored in the register \texttt{rax} at line~B0 (in red).
    A line with a darker color denotes a larger attention value. In an instruction, the first operand is destination and the second the source, and comment ``{\tt rax<-i}'' means that register {\tt rax} stores variable {\tt i}.
    }
    \label{fig:moti-attn}
\end{figure}

Fig.~\ref{fig:moti-attn}(a) shows part of the attention map of lines~9--12 in the code in Fig.~\ref{fig:moti-ex}(a),
by the CodeT5 model that works on source code.
Observe that with the help of variable name {\tt mean}, the model correctly correlates statements with program dependences, e.g., the strong attention between {\it mean} at line~11 and {\it mean} at line~9. 
In contrast, Fig.~\ref{fig:moti-attn}(b) shows the attention map of the corresponding binary code by 
the JTrans model~\cite{wang2022jtrans}. Here, {\tt rax}@A0 denotes variable {\tt i}, {\tt rax}@B0 denotes {\tt mean} and {\tt rax}@C0 denotes {\tt i}. Observe that the model has undesirable (strong) attention among the {\tt rax}'s.
Moreover, as the binaries for the programs in Fig.~\ref{fig:moti-ex}(a) and (c) 
are syntactically very similar.
The JTrans model produced highly similar (but problematic) attention maps for the two, causing its misclassification of the two as similar functions.
In Section~\ref{sec:eval:binsim}, our results show that \toolname outperforms the JTrans model on the binary similarity task by over 30\% in zero-shot settings.

\begin{figure}
    \centering
    \includegraphics[width=\linewidth]{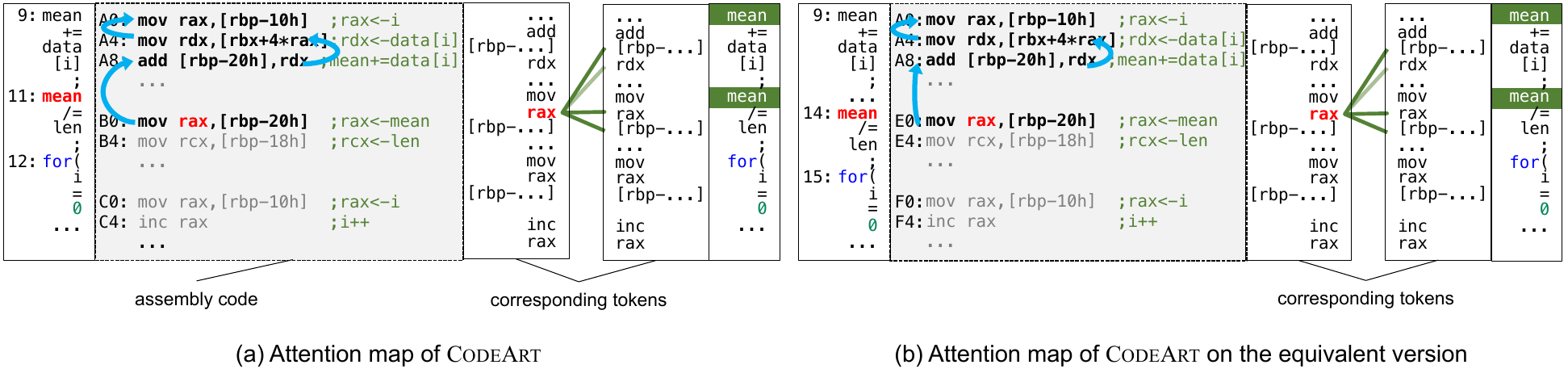}
    \caption{Attention maps by \toolname for variable {\tt mean} in the code snippets in Fig.~\ref{fig:moti-ex} (a) and (b). The instructions in bold are in the dependence context of the variable and the blue arrows denote dependences.
    }
    \label{fig:moti-attn-codeart}
\end{figure}

\smallskip
\noindent
{\bf GNN Models.}
GNNs
are also widely used in software engineering tasks~\cite{allamanis2017learning,li2019graph,bieber2020learning,guo2020graphspy,chen2021plur}.
The basic idea is that programs have explicit graph structures, such as control flow graph (CFG), abstract syntax tree (AST), and program dependence graph (PDG), which can be leveraged by GNNs. In GNNs, the embedding of a node  (e.g., a basic block in CFG) can be derived from the embeddings of neighboring nodes and the content of the node itself. 
They are hence a plausible solution
when symbols are missing.
However, as pointed out in ~\cite{wang2022jtrans,wu2021representing}, standard GNN based code models struggle to capture long-range dependences 
~\cite{zhu2020beyond}. Although  scaling GNNs in depth and width may mitigate the problem, it may in the meantime cause optimization instabilities and representation oversmoothing~\cite{li2018deeper,alon2020bottleneck,chen2020measuring}. 
Our experiments in Section~\ref{sec:eval:binsim} show that \toolname outperforms GNN based code models on the binary similarity task by over 20\%.

\smallskip
\noindent
{\bf Our Method.}
Our key observation is that when symbols are lost or not informative,  code becomes dissimilar to natural language products, and exhibits its own characteristics. First, {\em the semantics of a statement is the aggregation of all the statements that directly or transitively contribute to it and those that it contributes to.} For example, the meaning of variable {\tt mean} at line 11 in Fig.~\ref{fig:moti-ex}(a) is determined by lines 8-11 and lines 13-14, which form a context for {\tt mean}.
In an extreme case where all statements in a function are devoted to computing a final return value, the embedding of the value shall reflect the meaning of the whole function. This is quite different from how humans use natural languages. Second, {\em absolute code positions shall be de-emphasized when meaningful symbol information is not present to help disentangle interleaving contexts}. Program code tends to have interleaving contexts, which may not be a problem when symbols are present to help disentanglement. For example in Fig.~\ref{fig:moti-ex}(b), the context of {\tt mean} interleaves with the context for computing percentile, namely, lines 11-13. Although it may not be a problem when variable names are present to disentangle the contexts,
it becomes a lot more challenging when symbols are lost.

Based on the observations, we propose a new technique that is
based on Transformer,  and enhanced with a new pre-training method to explicitly regulate self-attention using masks.
In particular, it uses program analysis to determine possible dependences between instructions, including both data and control dependences. 
The masks ensure that a token can only pay attention to its {\em dependence context}, which includes the instructions that it directly or transitively depends on and those that are directly or transitively dependent on the token. As we will show in Section~\ref{sec:arcr}, additional masks are also derived to regulate attention in other types of contexts such as {\em instruction local contexts} that only allow an token to pay attention to other tokens within the same instruction and {\em global contexts} that model token co-occurrences (within and across instructions). The pre-training further helps the model learn which of these allowed attentions ought to be strong.
As shown in Fig.~\ref{fig:moti-attn-codeart}(a), 
the attention map by \toolname for the binary version of Fig.~\ref{fig:moti-ex}(a) closely resembles that when the source code is used (Fig.~\ref{fig:moti-attn}(a)). Moreover, the attention map by \toolname for the binary version of Fig.~\ref{fig:moti-ex}(b) (i.e., the reordered but equivalent version) is also highly similar (see Fig.~\ref{fig:moti-attn-codeart}(b)), despite their syntactic differences. In fact, \toolname produces similar attention maps for all three code snippets in Fig.~\ref{fig:moti-ex}. However, the different operations at line 11 of Fig.~\ref{fig:moti-ex} (a) and (c) cause different final embeddings, allowing
a binary similarity analysis built on our model to correctly recognize the similar functions.

\label{sec:moti:gcb}
\smallskip
\noindent
{\bf Comparison with GraphCodeBERT.} 
GraphCodeBERT~\cite{guo2020graphcodebert} is a source code representation model based on BERT architecture that leverages data-flow information in pre-training. By augmenting (appending) the [comment, source code] input with additional variable tokens, and forcing the variable tokens to align with their corresponding tokens in the source code part of the input, and to have the intended data-flow, GraphCodeBERT aims to have better embeddings.
Although GraphCodeBERT has demonstrated advantages over the vanilla CodeBERT~\cite{feng2020codebert}, porting it to handling binaries without symbols is challenging. In particular, its variable token alignment largely relies on the variable name equivalence. At the binary level, the multiple occurrences of a variable may have completely different register/memory tokens, rendering the alignment much more difficult. Without proper alignment, the data-flow training is infeasible.
This is supported by our evaluation results of a binary version of GraphCodeBERT in Section~\ref{sec:eval:gcb}.

\section{Design}

\begin{figure}
    \centering
    \includegraphics[width=\linewidth]{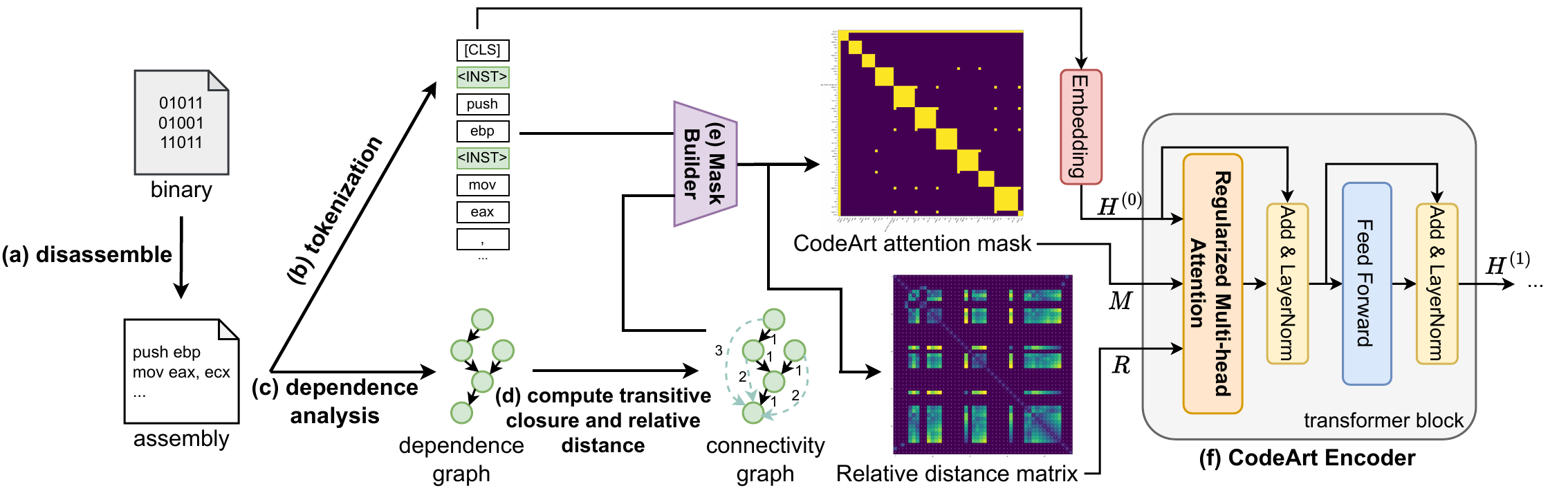}
    \caption{\toolname encoder pipeline
    }
    \label{fig:codeart-pipeline}
\end{figure}
\smallskip
\noindent
{\bf Overview.}
Fig.~\ref{fig:codeart-pipeline} shows the pipeline of \toolname encoder.
Given a binary executable, \toolname first disassembles the code (step~(a)). It then tokenizes the disassembly (step~(b)) and performs dependence analysis to derive both data and control dependences (step~(c)). 
In (d), the dependence transitive closures are computed for individual instructions, by traversing dependence edges in 
 both forward and backward directions. 
The closures are further transformed to connectivity graphs in which an edge is introduced between two nodes if one is reachable from the other. In \emph{Mask Builder} (e), the connectivity graph and the instruction tokens are leveraged to construct an attention mask and a relative distance matrix that measures the dependence distance between two connected nodes. They are further fed to a BERT-like Transformer model 
(step~(f))
to produce the output embedding. Our model is different from the vanilla BERT as it leverages the mask and the distance matrix to regulate attention.
The pre-training of \toolname model entails masking input tokens and masking/perturbing the connectivity graph.
In the following subsections, we introduce the individual components.

\subsection{Tokenization}
\label{sec:tokenization}

\toolname directly works on binary executables with symbols stripped. Given an executable, \toolname first disassembles it using IDA-Pro~\footnote{https://hex-rays.com/ida-pro/}.
The disassembled code is then tokenized. The tokenizer breaks each (x86-64)
instruction to multiple tokens. 
Specifically, opcode and operands are denoted by separate tokens. A compound operand may be further broken down to multiple tokens. 
For example, a memory read instruction {\tt mov rdx,[rbx+4*rax]} that reads a value from an address denoted by {\tt rbx+4*rax} to {\tt rdx} 
is tokenized to a sequence {\tt `mov', `rdx', `,', `[', `rbx', `+', `4', `*', `rax', `]'}.
In addition, we use a special delimiter token \texttt{<INST>} to denote the beginning of an instruction. This is critical to the attention regularization which we will discuss in Section~\ref{sec:arcr}. Formally, a tokenized sequence can be described as $A=\{\texttt{i}_1, a_{1, 1}, \cdots, a_{1,n_1},\texttt{i}_2, a_{2, 1},\cdots, a_{2, n_{2}}, \cdots \}$, where $i_t$ denotes the delimiter for the $t$-th instruction and $a_{t,n}$ the $n$-th token of the $t$-th instruction. Similar to other BERT-based models~\cite{devlin2018bert,liu2019roberta,feng2020codebert}, we prepend a \texttt{[CLS]} token to the token sequence. The final input sequence is hence $\boldsymbol{X}=[\texttt{[CLS]}, A]$.

\subsection{Dependence Analysis}
\label{sec:dep-analysis}

Given binary executables, \toolname first uses program analysis to determine both control dependences and data dependences between instructions, and then uses such information to regulate attention during training and inference.
In particular, \toolname employs IDA Pro~\cite{idapro} to construct the control flow graph (CFG) for each function. 
Using these CFGs, we resort to a conventional algorithm~\cite{ferrante1987program} to determine control dependences.
We additionally tailor a source-code data-flow analysis~\cite{allen1976program} to facilitate the analysis of data dependences in binaries.
This data-flow analysis begins by gathering a collection of variables accessed by each statement (or, in the context of binary analysis, an instruction) and subsequently determines the def-use relationship among these variables.
Precisely identifying variables accessed by a binary instruction is notably challenging~\cite{zhang2019bda}, given that all variables are compiled into plain registers and memory locations without any symbol information.
To this end, we adopt an approach that overestimates the memory regions an expression can potentially reference (and hence, the variables that an instruction with the expression can potentially access), shown as follows.
\begin{itemize}
    \item Expressions denoting stack memory addresses with statically decidable offsets (e.g., \texttt{[rsp + 0x20]}) are interpreted to merely reference the corresponding stack locations. 
    \item Expressions denoting stack memory addresses with statically undecidable offsets (e.g., \texttt{[rsp + rax * 8 + 0x30]} with {\tt rax} holding an input parameter)  are presumed to reference the entire stack frame of the corresponding function.
    \item Expressions pointing to addresses not statically associated with stack pointers (e.g., \texttt{[rbx + rax]} denoting some heap access) are conservatively considered to reference the entire memory space of the given binary.
\end{itemize}
For instance, given the assembly code: ``\texttt{(a) mov rax, rbx;} \texttt{(b) mov [rsp + 0x10], rax;} \texttt{(c) mov rcx, [rsp + 0x10];} \texttt{(d) mov rdx,  [rcx];}'', our analysis can derive three distinct data dependences from it, including \texttt{(a)$\leftarrow$(b)} through the register \texttt{rax}, \texttt{(b)$\leftarrow$(c)} involving the stack memory address \texttt{[rsp + 0x10]}, and  \texttt{(b)$\leftarrow$(d)} related to the non-stack memory address \texttt{[rcx]}. 
Note that \texttt{[rcx]} is assumed to possibly access the entire memory space, which encompasses \texttt{[rsp + 0x10]}.
It is worth noting that our dependence analysis is largely standard, and we include it for completeness. 
We do not claim contributions on the analysis.

\subsection{Model Architecture}
\label{sec:arch}

\begin{figure}
    \centering
    \includegraphics[width=0.7\linewidth]{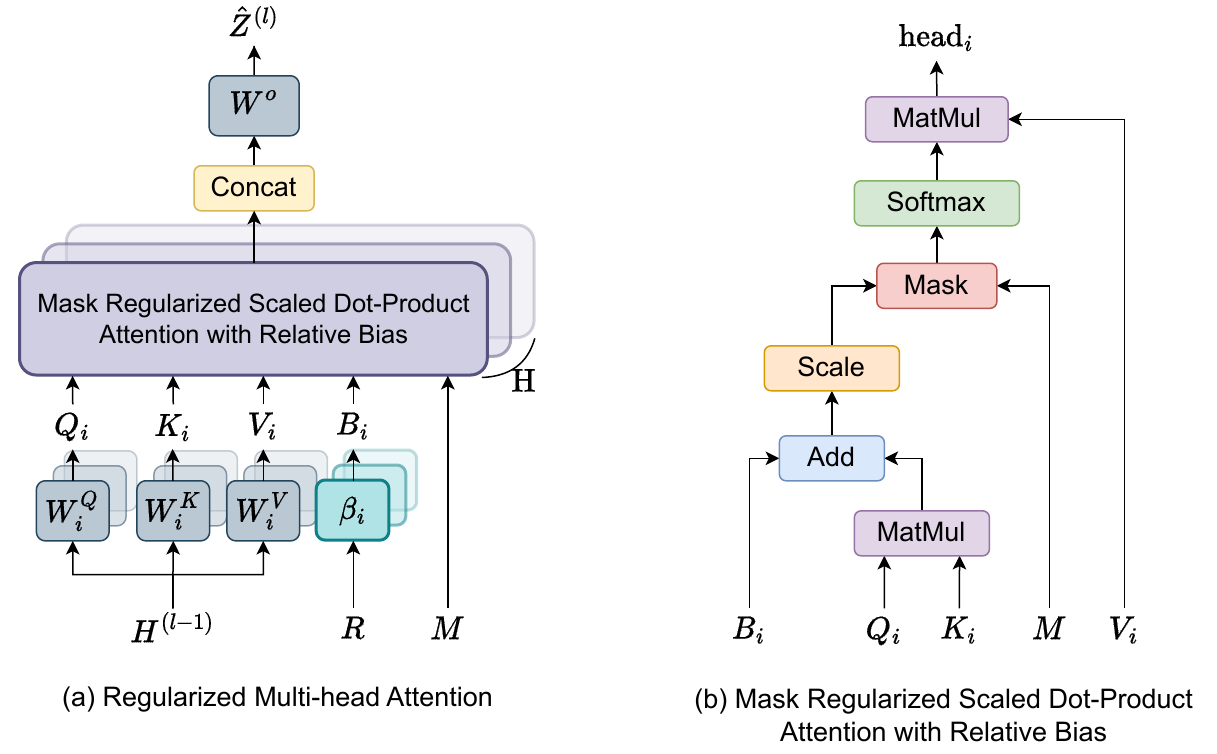}
    \caption{Regularized Multi-head Attention (with (b) illustrating a zoom-in view of part of (a)) %
    }
    \label{fig:attn-comp}
\end{figure}

Fig.~\ref{fig:codeart-pipeline}(f) shows our model architecture, which is a multi-layer Transformer encoder~\cite{vaswani2017attention}, with the component 
{\em Regularized Multi-head Attention} (RMA) containing the main differences from a standard BERT architecture~\cite{devlin2018bert,liu2019roberta}. %
As shown in the figure,
the encoder takes a sequence $\boldsymbol{X}$ of $N$ tokens and acquires a sequence of input vectors $\boldsymbol{H}^{(0)}=\left[\boldsymbol{h}_{1}^{(0)},\cdots,\boldsymbol{h}_{N}^{(0)}\right]\in\mathbb{R}^{N\times d_h}$ %
by summing the token embeddings and the corresponding trainable absolute position embeddings (the sum operation is not shown in the figure for brevity). Here, $d_h$ is the dimension of hidden states.
The output embedding
is then obtained by applying $L$ layers of transformer blocks,
each regulated by the \toolname attention mask $\boldsymbol{M}\in \mathbb{R}^{N\times N}$ and the relative distance matrix $\boldsymbol{R}\in \mathbb{N}^{N\times N}$, as exampled by the block in Fig.~\ref{fig:codeart-pipeline}(f). 
Formally, the hidden states after layer $l$ are denoted as $\boldsymbol{H}^{(l)}=\text{transformer\_block}\left(\boldsymbol{H}^{(l-1)}, \boldsymbol{M}, \boldsymbol{R}\right), \ l\in \{1..L\}$. 
A transformer block (Fig.~\ref{fig:codeart-pipeline} (f)) is formally defined as follows. 
\begin{equation}
\hat{\boldsymbol{Z}}^{(l)} = \text{RMA}(\boldsymbol{H}^{(l-1)}, \boldsymbol{M}, \boldsymbol{R}), \ \ \boldsymbol{Z}^{(l)} = \text{LN}(\hat{\boldsymbol{Z}}^{(l)} + \boldsymbol{H}^{(l-1)}), \ \  
\boldsymbol{H}^{(l)} = \text{LN}(\text{FFN}(\boldsymbol{Z}^{(l)}) + \boldsymbol{Z}^{(l)})
\end{equation}
Here RMA 
is an enhanced multi-head self-attention module~\cite{vaswani2017attention}, FFN is a two-layer feed-forward network, and LN is layer normalization. Their details are shown in Fig.~\ref{fig:attn-comp}. In particular, for each layer $l$, the output $\hat{Z}^{(l)}$ of RMA is computed as follows. We omit the layer annotation for simplicity.
\begin{equation}
\label{eq:qkv}
\boldsymbol{Q}_i = \boldsymbol{H}\boldsymbol{W}^Q_i,\ \ \boldsymbol{K}_i = \boldsymbol{H}\boldsymbol{W}_i^K, \ \ \boldsymbol{V}_i = \boldsymbol{H}\boldsymbol{W}_i^V
\end{equation}
\begin{equation}
\label{eq:scaled-dot-prod-attn}
   \text{head}_i = \text{softmax}\left(\frac{\boldsymbol{Q}_i\boldsymbol{K}_i^T + \boldsymbol{B}_i}{\sqrt{d_k}} + \boldsymbol{M}\right)\boldsymbol{V}_i
\end{equation}
\begin{equation}
\hat{\boldsymbol{Z}}=\left[\text{head}_1;\cdots;\text{head}_{\text{H}}\right]\boldsymbol{W}^o
\end{equation}

Here the $(l-1)$-th layer's output $\boldsymbol{H}^{(l-1)}\in\mathbb{R}^{N\times d_h}$ is converted to queries
$\boldsymbol{Q}_{i}=\left[\boldsymbol{q}_{i,1},\cdots,\boldsymbol{q}_{i,N}\right]$, keys $\boldsymbol{K}_{i}=\left[\boldsymbol{k}_{i,1},\cdots,\boldsymbol{k}_{i,N}\right]$ and values $\boldsymbol{V}_{i}=\left[\boldsymbol{v}_{i,1},\cdots,\boldsymbol{v}_{i,N}\right]$ for head $i$, by linear projections with weights $\boldsymbol{W}_i^Q$, $\boldsymbol{W}_i^K$, and $\boldsymbol{W}_i^V\in\mathbb{R}^{d_h\times d_k}$. $\text{H}$ is the number of heads, $d_k$ the head dimension, and $\boldsymbol{W}^o$ the weight for the linear projection after concatenating multiple head outputs.
$\boldsymbol{M}$ denotes a main difference between our architecture and vanilla Transformer models. It is the {\em multi-head attention mask}, where $\boldsymbol{M}_{uv}$ is set to either $0$ or $-\infty$ to determine whether token $u$ is allowed to attend to token $v$ or not, respectively. Intuitively, by setting parts of the mask to 0, we only allow the model to learn self-attention for the corresponding token pairs and preclude spurious attention.
We will further explain how to construct the mask later in the section.

As shown in Equation~\ref{eq:scaled-dot-prod-attn} and Fig.~\ref{fig:attn-comp}(b), another main difference between RMA and standard attention is that RMA incorporates relative positional embedding $\boldsymbol{B}_i$ to explicitly encode 
position relationship between  \texttt{<INST>} delimiter tokens (not between other tokens internal to an instruction).
Intuitively, it denotes the dependence distances between instructions as their absolute positions/distances could be misleading when symbols are missing.
The relative positional embedding is integrated in the form of bias in Equation~\ref{eq:scaled-dot-prod-attn}. 
It is computed as follows.
\begin{equation}
\label{eq:bias}
  \boldsymbol{B}_{i,uv} = \beta_i\left(\min(\boldsymbol{R}_{uv}, r_{\text{max}})\right)
\end{equation}
Here $\boldsymbol{R}$ is the relative distance matrix pre-computed by our analysis (details in the next section), and $\beta_i$ is the 
embedding function that maps a distance value to a trainable parameter for head $i$. The relative position relationship is learned through these parameters during training.  Variable $r_{\text{max}}$ is the max relative distance. Since larger relative distances are less frequent in practice, the model can hardly learn the relative bias for large relative distances. The max distance ensures that when the distance is too large, the bias becomes indistinguishable to that of at distance $r_{\text{max}}$. As such, the model understands that the distance is large and potentially out of distribution.

\subsection{Attention Regularization by Masking}
\label{sec:arcr}

\begin{figure}
    \centering
    \includegraphics[width=\linewidth]{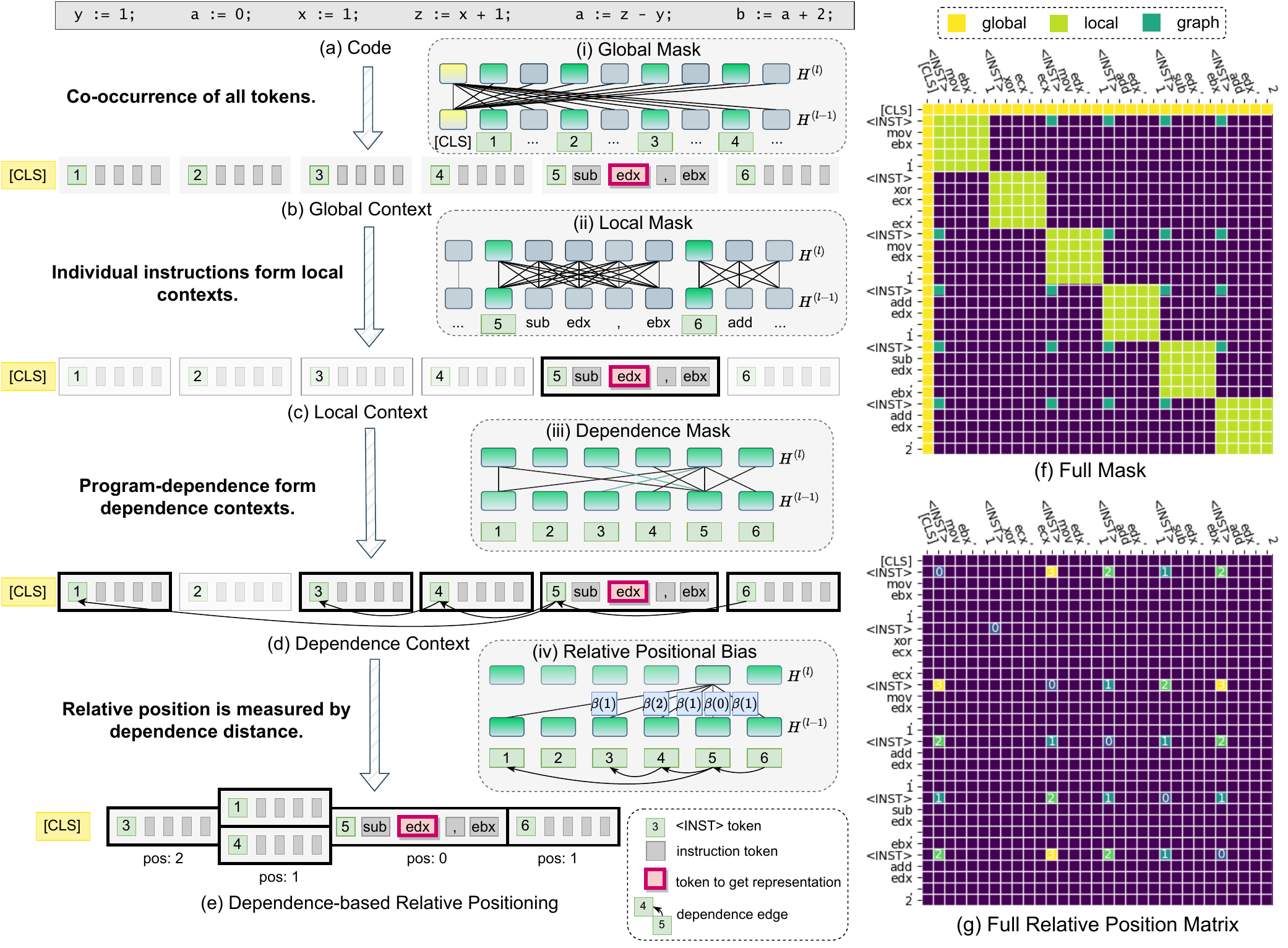}
    \caption{
    Contexts, attention masks and relative position matrix
    }
    \label{fig:ref-workflow}
\end{figure}

In this section,
we explain how we construct the attention mask $\boldsymbol{M}$.
As discussed earlier, without symbols, model training tends to determine contexts by names of primitive operands (e.g., register names) and absolute positions, which could be misleading. We leverage masking to direct the self-attention to the right places such that correct contexts can be extracted. 
Fig.~\ref{fig:ref-workflow} provides a 
conceptual illustration. 
In (a), a small program consisting of six instructions is shown in the first row (and in their source code form for readability), and each is broken down to multiple tokens. %
For example in the second row, the green blocks with numbers show the {\tt <INST>} tokens, each of which is followed by the (grey) tokens of the corresponding instruction. The entire sequence is preceded by a {\tt [CLS]} token.
Assume we are interested in the embedding of the second token of instruction 5 (the `{\tt edx}' token in red). 
First of all, we want the model to pay attention to 
the co-occurrences of all the tokens, without focusing on any specific ones. We call it the {\em global context}  (Fig.~\ref{fig:ref-workflow}(b)). 
Second, the model should pay attention to all the tokens in the same instruction as the red token when encoding it. We call it the {\em local context} of the red token (Fig.~\ref{fig:ref-workflow}(c)). 
Additionally, the model should pay attention to the sub-computation which the red token is in, namely, instructions \boxed{1}, \boxed{3}, \boxed{4}, and \boxed{6} 
 due to program dependences.  We call it the {\em dependence context} (Fig.~\ref{fig:ref-workflow}(d)).
Finally, the neighboring tokens of the red token should be measured by their dependence distances (i.e., how many dependence edges away) instead of their absolute positions.
We call it the {\em dependence based relative positioning} (Fig.~\ref{fig:ref-workflow}(e)). Intuitively, it rearranges the instructions by their dependence distances. For example, instruction 5 is closer to instructions 1, 4, 6 than instruction 3.

In standard Transformer models, the self-attention (without masking) allows a query $\boldsymbol{q}_u\in \mathbb{R}^{1\times d_k}$ (an element of $\boldsymbol{Q}$ in Equation~\ref{eq:qkv}, corresponding to a token) to attend to individual keys $\boldsymbol{k}_v\in \mathbb{R}^{1\times d_k}\ (v\in \{1..N\})$ (an element of $\boldsymbol{K}$ in Equation~\ref{eq:qkv}) by computing a scaled dot-product $\boldsymbol{q}_u\cdot \boldsymbol{k}_v^T/\sqrt{d_k}$ as the attention score, which is further normalized by a softmax function to get attention weights $o_{uv}$ for aggregating values $\boldsymbol{v}_v\in \mathbb{R}^{1\times d_k}\ (v\in \{1..N\})$
by
$\boldsymbol{z}_u=\sum_{v=1}^{N}o_{uv}\boldsymbol{v}_v$ (similar to Equation~\ref{eq:scaled-dot-prod-attn}).
The aforementioned
multi-step context extraction is achieved by {\em attention regularization} using masking. 
As shown in Equation~\ref{eq:scaled-dot-prod-attn}, an attention mask $\boldsymbol{M}$ is added to the dot-product before the softmax, which contains a 0 value (to enable attention) or a $-\infty$ value (to disable attention).
Note that adding $-\infty$ to the dot-product ensures that the attention weight $o_{uv}$ after the softmax becomes 0 so that the information cannot be aggregated from value $\boldsymbol{v}_v$ into the hidden state.
When the mask value is 0, the attention weight is the same as the default one, allowing self-attention.
$\boldsymbol{M}$ consists of three kinds of masks, corresponding to the aforementioned global, local, and dependence contexts. 

\smallskip
\noindent
{\bf Global Attention Mask $\boldsymbol{M}^{\text{Gl}}$.} As shown in Fig.~\ref{fig:ref-workflow}(i) in between the first and the second rows on the left,
the yellow \texttt{[CLS]} token 
is used to facilitate learning the global context.
Observe that it attends to all the tokens, and vice versa.
This 
allows the model to learn 
{\em co-occurrences}.
For example, the tokens in two separate instructions could indirectly attend to each other through the \texttt{[CLS]} token. The yellow cells in Fig.~\ref{fig:ref-workflow}(f) show the 0 values in $\boldsymbol{M}^{\text{Gl}}$ of our small program in the form of heat-map, with the legends the tokens and the black cells denoting $-\infty$.
Formally,

\begin{equation}
M^{\text{Gl}}_{i,j} = 
\begin{cases}
0, & X[i] = \texttt{[CLS]} \lor X[j] = \texttt{[CLS]},\\
-\infty, & \text{otherwise}.
\end{cases}
\end{equation}

\smallskip
\noindent
{\bf Local Attention Mask $\boldsymbol{M}^{\text{Lo}}$.}  $\boldsymbol{M}^{\text{Lo}}$ ensures a token inside an instruction has attention to all tokens in the same instruction.
Each instruction is broken down to an 
\texttt{<INST>} delimiter followed by a sequence of instruction tokens denoting opcode and operands (e.g., the 5th instruction in Fig.~\ref{fig:ref-workflow}(b)). 
As shown in Fig.~\ref{fig:ref-workflow}(ii),
each token in the 5th instruction attends to all the tokens within the instruction.
Moreover, any cross-instruction attention is forbidden in this mask. 
The green cells in Fig.~\ref{fig:ref-workflow}(f) show the 0 values in $\boldsymbol{M}^{\text{Lo}}$ of our small program. Formally, 

\begin{equation}
M^{\text{Lo}}_{i,j} =
\begin{cases}
0, & \texttt{instruction\_of}(X[i]) = \texttt{instruction\_of}(X[j]),\\
-\infty, & \text{otherwise}.
\end{cases}
\end{equation}

\smallskip
\noindent
{\bf Dependence Attention Mask $\boldsymbol{M}^{\text{Dep}}$.}
A straightforward method to construct the dependence mask is to directly reflect the directed program dependency graph $G_\text{dep}=(V, E_{\text{dep}})$ obtained by our dependency analysis through the \texttt{<INST>} tokens, namely, if an instruction $A$ is dependent on another instruction $B$ (i.e., there is a dependence edge $A\rightarrow B$, we allow $A$'s \texttt{<INST>} token to pay attention to $B$'s).
However, such a simple design suffers two problems: (1) 
one layer can only pass information from 1-hop neighbors such that signals from multi-hop neighbors become undesirably weak.
(2) the dependence attention is uni-directional whereas
bidirectional attention 
is proved to be better in Transformers pre-training~\cite{devlin2018bert}.

Thus, we use the Floyd-Warshall algorithm~\cite{cormen2022introduction} to transform the dependence graph to 
a connectivity graph $G_{\texttt{con}}=(V, E_{\text{con}}, \mathcal{D})$, where 
an undirected edge between two nodes denotes if one is reachable from the other in the original graph, and
$\mathcal{D}:V\times V\mapsto \mathbb{N}$ is the distance function which maps a connectivity edge to its path length in the original graph. 
Then, we construct the mask to enable bidirectional attention if a connectivity edge exists.
Fig.~\ref{fig:ref-workflow}(iii) 
shows the dependence mask. Observe that since instruction 5 is dependent on instructions 1 and 4, transitively dependent on 3, and 6 is dependent on 5, symmetric attention is allowed between 5 and 1, 3, 4, 5, and 6.
The blue cells in Fig.~\ref{fig:ref-workflow}(iii) show the 0 values in $\boldsymbol{M}^{\text{Dep}}$ of our small example. 
Note that although we only allow direct attention between {\tt <INST>} tokens, tokens internal to instructions attend to {\tt <INST>} and vice versa such that individual internal tokens can attend to other internal tokens in the dependence context through an additional layer of information propagation. Formally,

\begin{equation}
M^{\text{Dep}}_{i,j} =
\begin{cases}
0, & X[i] = \texttt{<INST>}
\land X[j] = \texttt{<INST>}\\
~ & 
\land 
(\texttt{instruction\_of}(X[i]) \ \textbf{trans-deps}\ \texttt{instruction\_of}(X[j]) \\
~ & \lor 
\texttt{instruction\_of}(X[j]) \ \textbf{trans-deps}\ \texttt{instruction\_of}(X[i])),\\
-\infty, & \text{otherwise}.
\end{cases}
\end{equation}

The final mask $\boldsymbol{M}$ is the union of above three masks, as in Fig.~\ref{fig:ref-workflow}(f).
The distance function $D$ of the connectivity graph is further used to construct the relative distance $\boldsymbol{R}$ in Equation~\ref{eq:bias}. 
Fig.~\ref{fig:ref-workflow}(iv) shows the relative positional biases for the 5th instruction (i.e., the annotations on attention edges).
Fig.~\ref{fig:ref-workflow}(g) shows the constructed $\boldsymbol{R}$ with different colors denoting different distances.

\subsection{Model Pre-training}
\label{sec:pretraining}

\begin{figure}
    \centering
    \includegraphics[width=\linewidth]{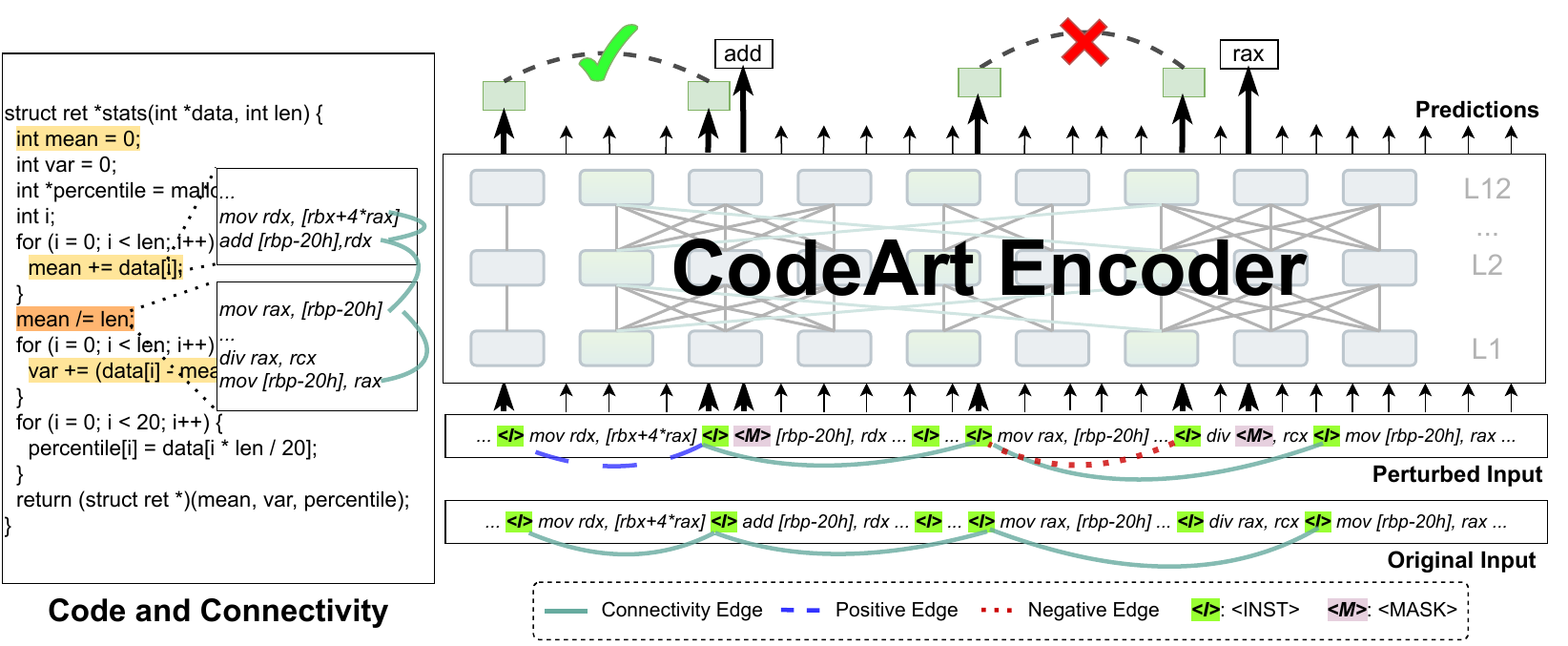}
    \caption{Pre-training in \toolname 
    }
    \label{fig:pretraining}
\end{figure}

The pre-training of \toolname consists of two parts: the traditional {\em Masked Language Modeling} (MLM) that masks parts of input tokens and {\em Masked Dependence Modeling} (MDM) that masks existing dependence edges and introduces spurious dependence edges.
The essence of MDM is that well-trained contextual embeddings shall be able to predict existing dependence edges (called {\em positive edges}) and preclude spurious edges (called {\em negative edges}).

\smallskip
\noindent
{\bf Masked Language Modeling.}
We perturb the input sequence $X$ following the MLM in RoBERTa~\cite{liu2019roberta}. In particular, we sample 15\% of the tokens in $X$. We then replace them with a \texttt{[MASK]} 80\% of time, with a random token 10\% of time, and have them unchanged in the remaining 10\%. During pre-training, the model is supposed to predict the masked tokens. Let $\tilde{X}$ be the perturbed sequence. the The MLM training loss is formally defined as follows.  
\begin{equation}
   \mathcal{L}_{\text{MLM}} = \sum_{x\in X^{\text{masked}}}-\log P\left(x \big| \tilde{X}, \tilde{\boldsymbol{M}}, \tilde{\boldsymbol{R}}\right)%
\end{equation}

where $\tilde{\boldsymbol{M}}$ and $\tilde{\boldsymbol{R}}$ are the perturbed attention mask and relative position matrix respectively, which we will discuss next.

\smallskip
\noindent
{\bf Masked Dependence Modeling.}
In each pre-training step, we randomly sample 40\% of the nodes in the connectivity graph $G_{\text{con}}=(V,E_{\text{con}})$. The sampled set is denoted as
$V_{s}\subset V$. 
Let $E_{\text{con}}(V_s)$ be the set of edges in the connectivity graph that have at least one node in $V_s$. It denotes the sampled positive edges.
We then sample an equal number of pair-wise relations in $V_{s}\times V \cup V \times V_{s}\ - \ E_{\text{con}}$, which denote the negative edges. 
These two edge sets form a balance set $E_b$ of positive and negative samples. We then force the model to learn to correctly classify the positive and negative edges during pre-training.
Let $\tilde{\boldsymbol{M}}$ and $\tilde{\boldsymbol{R}}$ be the perturbed attention mask and relative matrix after the sampled positive edges removed from $G_{\text{con}}$ and the negative edges added, and $\boldsymbol{h}_{u}$, $\boldsymbol{h}_{v}$  the hidden states of $u$ and $v$. The training loss is formally defined as follows. 
\begin{equation}
\mathcal{L}_{\text{EdgePred}} = \sum_{e_{uv}\in E_b}-\mathbb{I}(e_{uv}\in E_{\text{con}})\log p_{e_{uv}}-\mathbb{I}(e_{uv}\notin E_{\text{con}})\log(1-p_{e_{uv}})
\end{equation}
where
\begin{equation}
    p_{e_{uv}} = P\left(e_{uv}\big|
    \tilde{X}, 
    \tilde{\boldsymbol{M}}, \tilde{\boldsymbol{R}}
    \right) = \text{sigmoid}(\boldsymbol{h}_{u}\cdot\boldsymbol{h}_{v}^T)
\end{equation}
Note that even though MDM is similar to GraphCodeBERT's edge prediction and node alignment losses~\cite{guo2020graphcodebert} as they are all in the form of binary cross-entropy, there is a core difference that MDM predicts edges between \texttt{<INST>} tokens instead of variable tokens from original code as in GraphCodeBERT. This avoids the risk that edge prediction loss may affect the representation of variable tokens learned by the MLM loss in an undesirable way.

The final loss is the sum of $\mathcal{L}_{\text{MLM}}$ and $\mathcal{L}_{\text{EdgePred}}$.

\smallskip
\noindent
{\em Example.}
Fig.~\ref{fig:pretraining}
presents how we use the motivation example in pre-training. The left shows the source code with the sub-computation related to {\tt mean} highlighted. The x86 instructions corresponding to two of these statements are also shown. On the right, we show the model and the two types of masked modeling.
The original input sequence is shown at the bottom with the edges denoting connectivity (derived from dependences). %
The row above shows the perturbed sequence and the perturbed edges.
In particular, a positive edge (in blue) is selected 
and a negative edge (in red) is introduced. 
On top of the encoder, the output embeddings could be used to correctly classify the positive/negative edges.

\section{Experiment}
The implementation and pre-training details of \toolname can be found in Section~\ref{appendix:pretrain} in the supplementary material. 
We evaluate \toolname on three downstream tasks to demonstrate the effectiveness of the pre-training, including binary similarity, malware family classification, and type inference for binary executables.
We aim to answer the following research questions.

\begin{itemize}
    \item RQ1: How does \toolname perform on binary similarity analysis?
    \item RQ2: How does \toolname perform on malware family classification?
    \item RQ3: How does \toolname perform on binary type inference?
    \item RQ4: Model analysis and ablation study of \toolname.
    \item RQ5: How effective is \toolname compared to GraphCodeBERT-like pre-training?
\end{itemize}

\subsection{RQ1: Performance on Binary Similarity Analysis}
\label{sec:eval:binsim}
\smallskip\noindent
{\bf Setup.} Given an input binary function, a binary similarity analysis queries it among a pool of candidate functions, and tries to identify the function that is compiled from the same source code as the query function~\cite{wang2022jtrans}. It plays a critical role in many security related tasks, such as one day vulnerability detection~\cite{IMF,duan2020deepbindiff}, automatic software patching~\cite{shari2021automated}, and software plagiarism detection~\cite{luo2017semantics}. Machine-learning based binary similarity tools typically encode the query function and all the candidate functions to their embeddings. After that, the cosine similarity between the embeddings of the query function and each candidate function is computed. The candidate functions are then ranked by the similarity values. The function with the largest value is considered similar to the query function.

\smallskip\noindent
{\bf Dataset.} We use the BinaryCorp-3M dataset~\cite{wang2022jtrans} to finetune \toolname on the binary similarity task. It is the same training dataset used by SOTA transformer-based models~\cite{wang2022jtrans,xu2023diemph,zhu2023ktrans}. We use 7 real-world projects (i.e., Curl, Coreutils, Binutils, ImageMagick, SQLite, OpenSSL and Putty) as the test dataset. They are commonly used by the previous binary similarity works~\cite{marcelli2022machine,pei2020trex,IMF,xu2023pem}. 

\smallskip\noindent
{\bf Baseline.} We compare the performance of \toolname with the SOTA GNN-based model~\cite{marcelli2022machine}, {\em Graph Matching Network} (GMN)-based model~\cite{li2019graph}, 
and two SOTA Transformer-based models Jtrans~\cite{wang2022jtrans} and DiEmph~\cite{xu2023diemph}. GMN is a GNN-based model that takes as input a pair of programs and outputs a similarity value. It is worth noting that GMN does not generate embeddings for individual functions, and can only be used for pair-wise similarity analysis.
Note that although there are recent proposals that can achieve very good performance in binary similarity using advanced dynamic analysis such as~\cite{xu2023pem,IMF}, these techniques require executing the functions (using seed inputs). They are hence not directly comparable. 

\smallskip\noindent
{\bf Training Details.} For both \toolname and JTrans, we finetune their pre-trained models
on the BinaryCorp-3M dataset~\cite{wang2022jtrans}. Specifically, each training data sample is a triplet consisting of two binary functions compiled from a same source code function, and another binary function compiled from a different source code function. The training loss is the triplet loss enforcing a large cosine similarity between similar function pairs and a small similarity between dissimilar ones. For DiEmph, the GNN-based model, and the GMN-based model, we follow the scripts provided by the authors~\cite{marcelli2022machine,xu2023diemph} to train the models.
For all models, we use the Coreutils project as the validation dataset to select best checkpoints, and report the performance on the remaining 6 projects. 

\smallskip\noindent
{\bf Metrics.} Following previous work~\cite{wang2022jtrans}, we use recall@1 as the metric to evaluate a binary similarity model. Specifically, suppose that we make $N$ queries, recall@1 is computed as $N$ divided by the number of queries that the function compiled from the same source code is correctly returned as the most similar function. We also adapt the setup of previous work to evaluate a binary similarity model with different sizes of candidate function pools. Intuitively, a larger pool size means a more challenging setup for a model.

\begin{figure}[t]
    \centering
    \includegraphics[width=0.95\linewidth]{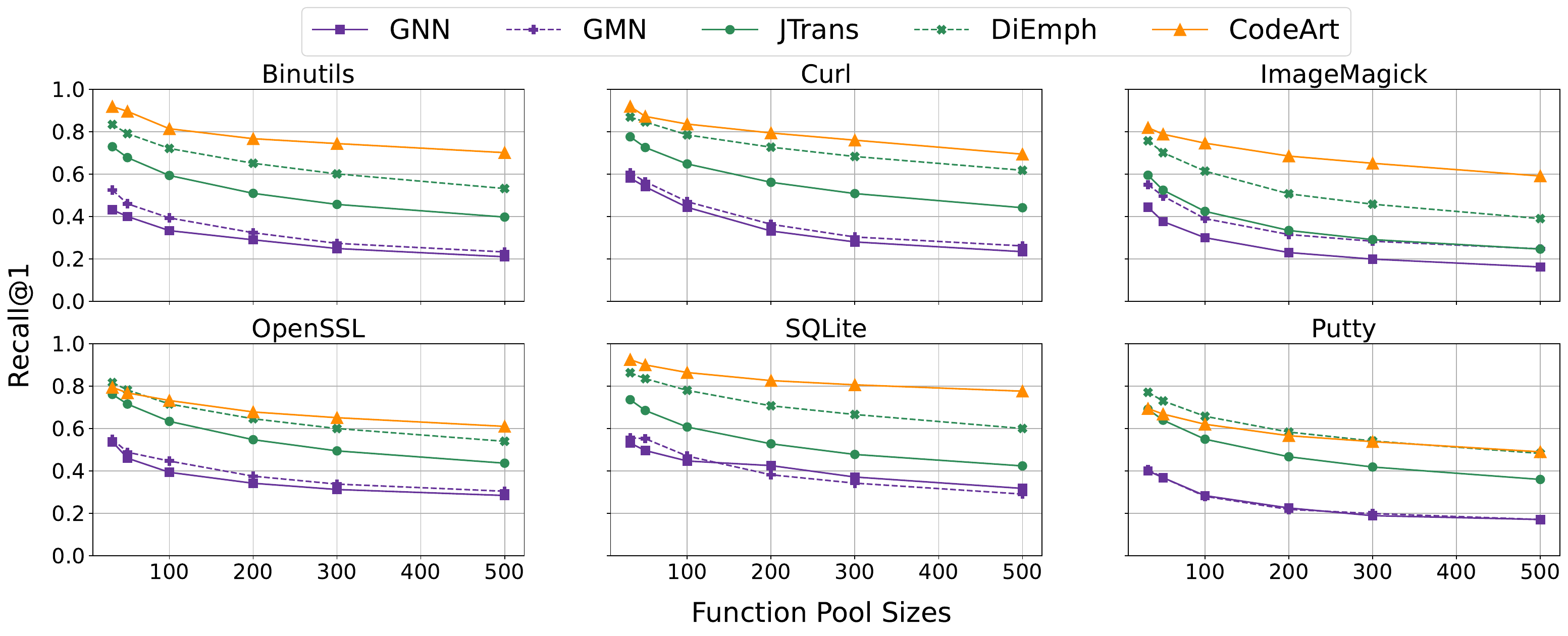}
    \caption{Performance (recall@1) of \toolname compared to the SOTA GNN-based model, GMN-based model and Transformer-based models. The x-axes denote different sizes of candidate function pools, and the y-axes denote the performance in terms of recall@1.}
    \label{fig:eval-binsim-perf}
\end{figure}

\smallskip\noindent
{\bf Results.} The results are shown in Fig.~\ref{fig:eval-binsim-perf}. We can see that \toolname outperforms the GNN-based model and the GMN-based model by a large margin in all setups. 
The improvement is largely due to Transformer models' better capability of capturing long-range dependences in a data-rich scenario, compared to GNNs. %
For most projects, \toolname significantly outperforms the Transformer-based models in challenging setups (i.e., a pool size larger than 100). The improvement demonstrates that \toolname is able to encode program semantics more precisely. For Putty, \toolname achieves comparable performance to the previous SOTA model DiEmph. We investigate the cases and found functions in Putty are shorter compared to other projects. For example, more than 75\% functions in Putty have less than 50 instructions, while, on the other hand, more than half of the functions in Binutils are longer than that. For those relatively simple functions, the problem of spurious correlations (caused by the lack of symbols) in the baseline models is not as severe and thus the improvement introduced by \toolname is not significant. Also, in simpler setups (i.e., a pool size smaller than 100), both DiEmph and \toolname can achieve good performance. 

To conclude, the results on the binary similarity task indicate \toolname can generate embeddings that encode program semantics more precisely, benefiting more realistic use scenarios\cite{wang2022jtrans}.

\smallskip\noindent
{\bf Zero-Shot Performance.} 
The performance on the binary similarity task relies heavily on the quality of function embeddings generated by a pre-trained model. We thus further evaluate \toolname on the binary similarity task with the zero-shot setup to measure the effectiveness of pre-training. Specifically, we directly use the embeddings generated by the pre-trained \toolname model without finetuning it on the binary similarity task. We compare its performance to the pre-trained JTrans model, which is pre-trained on the same dataset as \toolname. The results are shown in Table~\ref{tab:eval:zeroshot}.
We can see that \toolname demonstrates strong zero-shot performance by achieving over 30\% higher performance on all setups. It indicates that the function-level semantics are learned by \toolname during the pre-training process. We also conduct $t$-test between \toolname and JTrans' performance and results show that the improvement is statistically significant with $p$-values 1e-6, 6e-8, 1e-6, 2e-9, 3e-9 for pool size 32, 50, 100, 200, 500 respectively.

\begin{table*}[t!]
\centering
\footnotesize
\setlength{\tabcolsep}{4pt}
\caption{
Zero-Shot performance (recall@1) of \toolname compared to JTrans. 
Each row lists performance of both models w.r.t. one binary project with different sizes of candidate function pools.
}
\label{tab:eval:zeroshot}
\centering
\begin{tabular}{rcgcgcgcgcgcg}
\toprule
    \multirow{2.5}{*}{Programs} & \multicolumn{2}{c}{Pool size 32} & \multicolumn{2}{c}{Pool size 50} & \multicolumn{2}{c}{Pool size 100} & \multicolumn{2}{c}{Pool size 200} & \multicolumn{2}{c}{Pool size 500} \\
         
     \cmidrule(lr){2-3} \cmidrule(lr){4-5} 
     \cmidrule(lr){6-7} \cmidrule(lr){8-9}
     \cmidrule(lr){10-11}
     & JTrans & \cellcolor{white}\toolname & JTrans & \cellcolor{white}\toolname  & JTrans & \cellcolor{white}\toolname  & JTrans & \cellcolor{white}\toolname  & JTrans & \cellcolor{white}\toolname \\
     \midrule

Binutils & 0.30 &	0.71 &	0.26 &	0.68 &	0.20 &	0.60 &	0.14 &	0.57 &	0.09 &	0.48 \\
Curl & 0.32 &	0.71 &	0.28 &	0.66 &	0.21 &	0.63 &	0.16 &	0.56 &	0.12 &	0.47 \\
ImageMagick & 0.28 &	0.60 &	0.23 &	0.55 &	0.18 &	0.49 &	0.13 &	0.45 &	0.08 &	0.38 \\
OpenSSL & 0.33 &	0.61 &	0.28 &	0.55 &	0.22 &	0.52 &	0.17 &	0.49 &	0.12 &	0.43 \\
SQLite & 0.25 &	0.75 &	0.21 &	0.76 &	0.15 &	0.69 &	0.11 &	0.62 &	0.07 &	0.55 \\
Putty & 0.30 &	0.49 &	0.25 &	0.45 &	0.19 &	0.41 &	0.14 &	0.36 &	0.09 &	0.31 \\
\midrule
\textbf{Average }& \textbf{0.30} &	\textbf{0.64} &	\textbf{0.25} &	\textbf{0.61} &	\textbf{0.19} &	\textbf{0.56} &	\textbf{0.14} &	\textbf{0.51} &	\textbf{0.10} &	\textbf{0.44} \\

\bottomrule
\end{tabular}
\vspace{3pt}
\end{table*}

\subsection{RQ2: Performance on Malware Family Classification}

\smallskip\noindent
{\bf Setup.} To bypass malware detectors, the authors of a malware may create a set of variants of the malware by changing the code while keeping the malicious logic. Such a set of variants is often referred to as a malware family. Malware family classification aims to attribute a given malware sample to a known family. It can help many downstream security applications such as malware detection~\cite{xu2021mff} and malware authorship attribution~\cite{gray2021identifying}.

In many cases, it is very hard, if not impossible, to precisely attribute a malware sample to a single family. In our dataset, a malware sample may have multiple labels.
We therefore use a multi-label classification setting which performs independent binary classification on each class.

\smallskip
\noindent
{\bf Dataset.}
We build a dataset with the malware samples and labels provided by VirusTotal~\footnote{https://www.virustotal.com} and VirusShare~\footnote{https://virusshare.com}.
Since real-world malware families follow a long tail distribution, we use the most frequent 100 families to build our classification dataset. It consists of 5484 malware samples. We randomly split the dataset into train, validation, and test set with a ratio of 70\%, 10\%, and 20\%. 

\smallskip
\noindent
{\bf Baseline.} To demonstrate the effectiveness of \toolname, we choose JTrans as a Transformer-based baseline because it is pre-trained on the same dataset~\cite{wang2022jtrans} as \toolname. Moreover, a mainstream of malware family classification leverages Convolutional Neural Network (CNN) to extract features from malware samples~\cite{nataraj2011malware,ma2021comprehensive}. Due to the lack of available code artifacts, we additionally re-implement a CNN SOTA~\cite{ma2021comprehensive} based on ResNet-50 and use it as another baseline.

\smallskip
\noindent
{\bf Metrics.}
We use three metrics typically used in multi-label classification tasks to measure the performance, i.e., ROC-AUC score, Label Ranking Average Precision score (LRAP), and Label Ranking Loss (LRL)~\cite{tsoumakas2010mining}. ROC-AUC score is the averaged area under the ROC curve, where ROC curve is the plot of the true positive rate (TPR) against the false positive rate (FPR), at various thresholds. LRAP and LRL rank the predicted probability of all possible types, and compute the scores based on the rank of the groundtruth type. Intuitively, a higher rank of the ground-truth type indicates the model has better capability of identifying the correct type of the input code.
 Please refer to Section~\ref{appendix:mfc} in the supplementary material for details.

\smallskip
\noindent
{\bf Training Details.}
\toolname is a function-level encoder. To encode a whole binary to an embedding,
we encode the forefront $k$ functions separately by our encoder and use an attention pooling strategy~\cite{luong2015effective} to pool the resulting $k$ embeddings into a single binary-level embedding. With the binary-level embedding, we stack a standard classification head to do the prediction. 
We similarly implement a baseline classifier for JTrans.
For the CNN-based baseline, we follow %
~\cite{ma2021comprehensive} to encode a malware binary to an image. Given a malware binary, we encode the entire binary byte-by-byte to a grey-scale image, compress it to a size of 256$\times$256, and convert it according to ResNet-50's input format. 
Then we train a ResNet-50 classifier on the images encoded from malware samples.

\smallskip
\noindent
{\bf Results.} We report the averaged results on 100 classes in Table~\ref{tab:mfc}. Columns 3--5 list the three metrics. Each row presents the performance of one model/setup. Rows 1--2, 3--4, 5--6 present the related models when encoding the forefront 2,3,4 functions, respectively. The last row presents the CNN result, which is on whole binaries. We can see that  \toolname achieves the best performance when encoding the forefront 4 functions. It is worth noting that \toolname achieves good performance when encoding even only 2 functions, better than CNN and JTrans.
The performance of JTrans degrades when the number of encoded functions decreases because encoding fewer functions requires a model to understand individual functions more precisely.
The improvement of \toolname compared to JTrans is significant with $p$-values 3e-4, 2e-4, 3e-3 for AUC-ROC, LRAP, LRL respectively.
These results
demonstrate that the functional semantics learned in the pre-training of \toolname is more informative for malware family classification.
Detailed (per-class) results can be found in Section~\ref{appendix:mfc} in the supplementary material. %

\begin{table}[t]
\begin{minipage}[t]{.4\linewidth}
    \centering
    \scriptsize
    \setlength{\tabcolsep}{1.5pt}
    \caption{Malware Family Classification  %
    }
    \begin{tabular}{rrccc}
    \toprule
     & & AUC ($\uparrow$) & LRAP ($\uparrow$) & LRL ($\downarrow$) \\
     \midrule
     \multirow{2}{*}{2 Funcs.} & JTrans & 0.705 & 0.205 & 0.300 \\
                                 & \toolname & 0.925 & \textbf{0.597} & 0.086 \\
                                 \midrule
    \multirow{2}{*}{3 Funcs.} & JTrans  & 0.898 & 0.478 & 0.111 \\
                                 & \toolname & 0.924 & \textbf{0.597} & 0.086 \\
                                 \midrule
    \multirow{2}{*}{4 Funcs.} & JTrans & 0.901 & 0.489 & 0.107 \\
                                 & \toolname & \textbf{0.928} & \textbf{0.597} & \textbf{0.084} \\
                                 \midrule
                                & CNN  & 0.834 & 0.326 & 0.183 \\

    \bottomrule
    \end{tabular}    
    \label{tab:mfc}
\end{minipage}
\hspace{.03\linewidth}
\begin{minipage}[t]{.45\linewidth}
    
    \centering
    \scriptsize
    \setlength{\tabcolsep}{1.5pt}
    \caption{Ablation Study Using Zero-Shot Binary Similarity with Pool-Size 100 on Coreutils
    }
    \begin{tabular}{lcccc}
        \toprule
        ~ & Recall@1 & Recall@3 & Recall@5 &  MRR \\
        \midrule
        \toolname-3M  &  0.394 & 0.548 & 0.604 &  0.486 \\
        \midrule
        \quad w/o local mask &  0.030 & 0.096 & 0.138	&  0.082\\
        \quad w/o trans-closure & 0.346 & 0.498 & 0.590 & 0.447\\
        \quad max-trans-closure 4 & 0.322 & 0.504 & 0.574 &  0.429\\
        \quad max-trans-closure 6 & 0.298 & 0.456 & 0.562 &  0.406\\
        \quad w/o rel-pos-bias & 0.226 & 0.376 & 0.464 & 0.335\\ 
        \bottomrule
    \end{tabular}
    \label{tab:ablation}
\end{minipage}
\end{table}

\subsection{RQ3: Performance on Type Inference}

\smallskip\noindent
{\bf Setup.} Binary programs do not have high-level type information as in the source code (e.g., pointer, struct, array). The binary type inference task aims to recover the high-level type information from assembly code. Specifically, given a location (e.g., a register) in the binary code, a type inference tool outputs the possible high-level type corresponding to the location. It helps a reverse-engineer understand binary code. Moreover, the recovered type information is often an input for other analysis such as vulnerability detection~\cite{lin2010automatic}, decompilation~\footnote{https://hex-rays.com/ida-pro/}, and legacy code hardening\cite{fawaz2015anatomization}.
Machine-learning based binary type inference tools typically formulate the type inference problem as a sequence labeling task~\cite{pei2021stateformer,chen2022augmenting}.
Following the setup of StateFormer~\cite{pei2021stateformer}, a SOTA binary type inference model, we define 35 common types as labels and include a special label \texttt{no-access} for tokens without groundtruth types.

\smallskip
\noindent
{\bf Dataset.}
We reuse the dataset in StateFormer's repo~\footnote{https://github.com/CUMLSec/stateformer/}. As our model is only for x86-64, we only perform experiments on the x86-64 part although the original dataset includes multiple architectures. We randomly split the dataset into train, validation, and test sets with a ratio of 85\%, 5\%, and 10\%.

\smallskip
\noindent
{\bf Baseline.}
We directly compare to the reported results of StateFormer, which was
finetuned on the same type inference dataset. 

\smallskip
\noindent
{\bf Training Details.}
We finetune \toolname on the training set by stacking a 1-layer linear classification head to project the last-layer hidden states of tokens to class probabilities. We choose the best-performing checkpoint on the validation set and report the results on the test set.

\smallskip
\noindent
{\bf Metrics.}
We follow the original setting of using precision, recall, and F1-score as the metrics. Here, precision $P=TP/(TP + FP)$, recall $R=TP/(TP + FN)$, and $F1=2\times P \times R / (P + R)$, where $TP$ is the number of correctly predicted types, $FP$ is the number of incorrectly predicted types, and $FN$ is the number of types incorrectly predicted as  \texttt{no-access}.%

\begin{figure}
    \centering 
    \includegraphics[width=.8\linewidth]{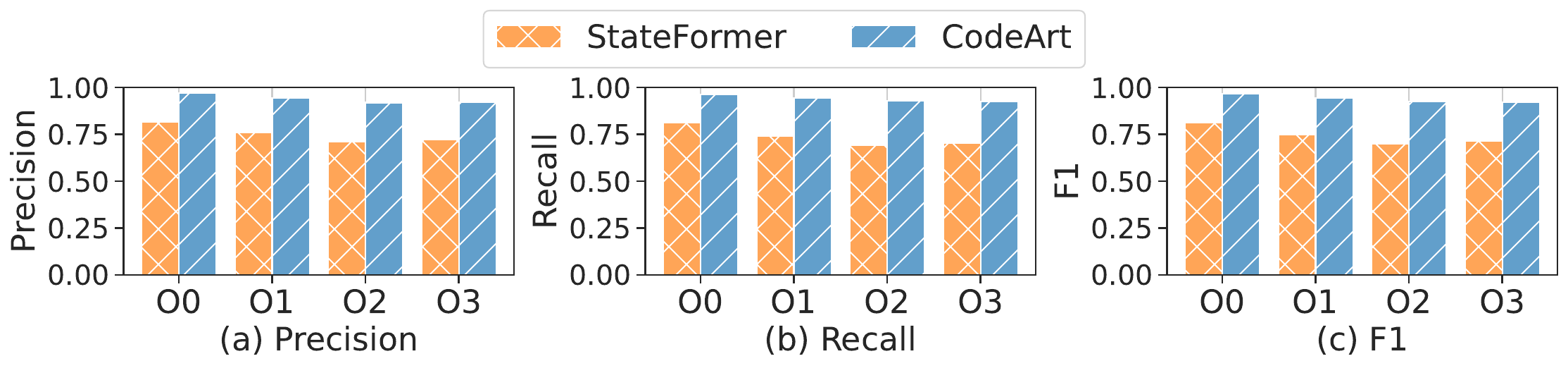}
    \caption{Binary Type Inference Results on Different Optimization Levels
    }
    \vspace{-10pt}
    \label{fig:ca_sf_optm}
\end{figure}

\noindent
{\bf Results on Different Optimization Levels.} 
We show the results of \toolname compared to StateFormer in Fig.~\ref{fig:ca_sf_optm}. The three sub-figures show the performance of both models in terms of precision, recall, and F1, respectively. The x-axes denote the optimization levels. We can see that \toolname achieves better performance in all 
optimization levels significantly with $p$-values 2e-8, 3e-8, and 3e-8 for precision, recall, and F1-score respectively. We attribute the improvement to \toolname's capability of precisely encoding tokens in their dependence contexts. Moreover, the performance of \toolname is more stable across different optimizations, suggesting \toolname is more robust to syntactic code changes %
introduced by compiler optimizations.

\subsection{RQ4: Model Analysis and Ablation Studies on What Helps \toolname}

\smallskip\noindent
{\bf Model Analysis.}
We now discuss \toolname in detail. Architecture-wise, \toolname largely aligns with the BERT-based models in hyperparameters and has slightly more trainable parameters 
(1296 in addition to 109 million) due to the additional parameters for relative positional bias. The average time of mask construction for a single sample is about 2.5ms. Note that %
mask construction can be done in parallel with the forwarding and back-propagation of the model.
We leave the optimization to our future work. %
The pre-training dataset is the BinaryCorp-26M dataset~\cite{wang2022jtrans}. We pre-train \toolname on the dataset for 25k steps. For more pre-training details, please refer to Section~\ref{appendix:pretrain} in the supplementary material. 

\smallskip\noindent
{\bf Ablation Studies.}
To study how each component in \toolname contributes to the final performance, we alter the pre-training process with different setups and observe how the performance changes. As detailed in Section~\ref{sec:eval:binsim}, the zero-shot performance on the binary similarity task reflects the quality of embeddings generated by a model and thus demonstrates the effectiveness of pre-training. Therefore, we leverage the zero-shot binary similarity performance on Coreutils (with a pool size of 100) as the metric for pre-training.
Due to the limitation of computation resources, we conduct the ablation study on the BinaryCorp-3M dataset. Note that all other evaluations are conducted on the \toolname model pre-trained on the larger BinaryCorp-26M dataset.

\begin{table}[]
    \centering
    \scriptsize
    \caption{Ablation Study on Other Downstream Tasks that Require Fine-tuning
    }
    \begin{tabular}{lccccccc}
        \toprule
        ~ & MFC-AUC & MFC-LRAP & MFC-LRL &  TI-PR & TI-Recall & TI-F1 \\
        \midrule
        CodeArt-3M  &  0.89	& 0.49 &    0.10   &        0.96 &	0.96 &	0.96\\
        \midrule
        w/o local mask &  0.86	& 0.40 &	0.13	& 0.92 &	0.91 &	0.91 \\
        w/o trans-closure & 0.86	& 0.47 &	0.14    & 0.94 &	0.93 &	0.94 \\
        max-trans-closure 4 & 0.90 &	0.50 &	0.10    & 0.96 &	0.96 &	0.96 \\
        max-trans-closure 6 & 0.90 &	0.47 &	0.11    & 0.96 &	0.95 &	0.95\\
        w/o rel-pos-bias & 0.87	& 0.47 &	0.12    & 0.96 &	0.96 &	0.96\\ 
        \bottomrule
    \end{tabular}
    \label{tab:ablation2}
\end{table}

The results are in Table~\ref{tab:ablation}. Each row denotes a variant of \toolname and the first row lists the performance achieved with the default setup (denoted by \toolname-3M).
We first remove the \textit{local attention masks} from \toolname (denoted by w/o local mask). Without the local masks, \toolname  degenerates to a pure masked language model, since all tokens can freely attend to each other and 
there is no mechanism to ensure that \texttt{<INST>} nodes are aligned with the corresponding instructions, rendering the dependence modeling ineffective. 
Hence, the performance dramatically drops. 

To demonstrate the necessity of 
modeling
transitive dependences  
in \toolname,
we implement a variant that does not include the \textit{transitive dependence closure} when constructing the dependence attention masks (denoted by w/o trans-closure). 
That is, the variant mimics the behavior of neighborhood-local aggregating GNNs, meaning that an instruction can only attend to its direct neighbours in the dependence graph. It 
weakens the perception of transitive dependences. 
Observe that the performance degrades by 4.8\%. It demonstrates that computing transitive closures and using  connectivity graphs indeed helps.

We further limit the maximum dependence distances when computing the transitive closures. Specifically, max-trans-closure 4 and max-trans-closure 6 denote variants that only include nodes reachable within 4 and 6 dependence edges,
respectively. We can see that the default \toolname (which does not limit the maximum distance) surpasses the two variants.

Moreover, we remove the \textit{relative positional bias}
from \toolname (denoted by w/o rel-pos-bias). That is, for a given instruction, all the instructions in its dependence context are considered having the same distance to it. We can see that the recall@1 degrades by 17\%. It validates that the relative position bias indeed helps the model distinguish instructions with different distances,
enabling the model to learn more precise semantics.

In addition to zero-shot binary similarity, we present the ablation study results for other tasks that require fine-tuning in Table~\ref{tab:ablation2}, where ``MFC'' denotes ``Malware Family Classification'' and ``TI'' denotes ``Type Inference''. We can see that \toolname's default setting demonstrates best overall performance, indicating its strong generalizability on all tasks.

\subsection{RQ5: Comparison with GraphCodeBERT-like Pre-training}
\label{sec:eval:gcb}
As discussed in Section~\ref{sec:moti:gcb}, GraphCodeBERT (GCB) leverages data-flow information 
to enhance the pre-training process. 
We compare the pre-training of \toolname to GCB's in terms of {\em optimization stability} and {\em zero-shot performance on the binary similarity task}.

\smallskip\noindent
{\bf Training Details.} Since the GCB pre-training code  is not publicly available and the origin GCB only worked on source code, we implement a pre-training pipeline following GCB's design and use BinaryCorp-3M\cite{wang2022jtrans} to pre-train it for the binary code. 
We detail how we implement the loss of GCB on binary code in Section~\ref{appendix:gcb-like} in the supplementary material.
Specifically, we implement two variants of the training pipeline: (1)~{\em GCB-like-default} that faithfully follows the loss design in GCB. (2)~{\em GCB-like-weaker-reg} that makes the additional losses smaller which means weaker regularization for the MLM loss.

\smallskip\noindent
{\bf Optimization Stability.} 
\toolname has much better stability.
The training curves of both \toolname and GraphCodeBERT-like pre-training are shown in Fig.~\ref{fig:gcb-curve} in Section~\ref{appendix:gcb-like} in the supplementary material.

\smallskip\noindent
{\bf Performance Difference.} As is shown in Table~\ref{tab:compare-gcb}, {\em GCB-like-weaker-reg} has better performance than {\em GCB-like-default} due to its stabler optimization. However, both variants are not comparable to \toolname, which demonstrates the effectiveness of our approach when symbols are lacking.

\begin{table}[t]
    \centering
    \footnotesize
    \caption{GrapCodeBERT-like Model Results on Zero-Shot Coreutils Binary Similarity with Pool-Size 100}
    \begin{tabular}{lccccc}
        \toprule
        ~ & Recall@1 & Recall@3 & Recall@5 & Recall@10 & MRR \\
        \midrule
        \toolname -3M  &  0.394 & 0.548 & 0.604 & 0.696 & 0.486 \\
        \midrule
        \quad GCB-like-default & 0.030 & 0.092 & 0.152 & 0.264 & 0.085 \\
        \quad GCB-like-weaker-reg & 0.082 & 0.176 & 0.266 & 0.394 & 0.159\\
        \bottomrule
    \end{tabular}
    \label{tab:compare-gcb}
\end{table}

\section{Related Work}
\noindent
{\bf Language Models for Code. }
A large body of work focuses on pre-training transformer-based language models on source code~\cite{kanade2020learning,feng2020codebert,ding2023traced,ahmad2021unified,wang-etal-2021-codet5}. They typically perceive source code as a sequence of tokens and utilize the rich natural language artifacts in software, such as symbol names and code comments, to help models understand code semantics. Due to the intrinsic structure of code, researchers have proposed to enhance pre-training with structural information, such as Abstract Syntax Trees (ASTs)~\cite{guo2022unixcoder}. A few works further explore ways to leverage program semantics to improve the quality of language models.
For example, OSCAR~\cite{peng2021could} enhances an IR-level code model with operational semantics of programs. They augment model inputs with abstract program states obtained from static analysis.
Their efforts are orthogonal to ours. 
GraphCodeBERT~\cite{guo2020graphcodebert} enhances MLM with data-flow graph structure, and we provide a detailed comparison in Section~\ref{sec:eval:gcb}. Unlike all the structure-enhanced approaches, our initiative is the first to understand the code structure from a language perspective. Our encoding pipeline and pre-training methods explicitly model the code language characteristics.

Additionally, there are some transformer-based models on binary code designed for specific downstream tasks~\cite{pei2020trex,wang2022jtrans,pei2021stateformer,jin2022symlm}.
In contrast, \toolname is designed to pre-train a general model that supports various downstream tasks.

\smallskip
\noindent
{\bf Neural Networks for Explicit Code Structure. } 
Apart from transformer-based language models, numerous code models explicitly encode code structures for specific software engineering tasks, such as program differencing~\cite{gao2024sigmadiff}, aligning code across platforms~\cite{kim2022improving}, disassembling binary code~\cite{yu2022deepdi}, software maintenance~\cite{dong2022fira}, and code completion~\cite{sun2020treegen}.
While these techniques leverage graph structures to derive embeddings, they usually require supervised training.
Our empirical results demonstrate that integrating self-supervised learning and self-attention in Transformers with dependence graphs is not only feasible but can also lead to superior performance.  
Moreover, adapting these models for pre-training tasks requires non-trivial engineering efforts~\cite{li2018deeper,alon2020bottleneck,chen2020measuring}. In contrast, \toolname proposes a unique attention regularization method that is fully compatible with existing efficient and scalable implementations for NLP Transformers. Thus, the pre-training of \toolname can easily scale to larger datasets.

\smallskip
\noindent
{\bf Binary Program Analysis. }
\toolname works at the level of assembly code, leveraging 
disassemblers~\cite{yu2022deepdi} to decode binary files into textual assembly code. There exists a body of research focusing on binary program dependence analysis~\cite{zhang2019bda}, which could enhance the analysis components in \toolname. 
Our work is built upon
these fundamental binary program analyses.

\section{Conclusion}

We introduce a novel method for pre-training Transformer-based code models in scenarios where symbols are lacking.
This method features an innovative attention regularization technique that leverages program analysis to derive potential dependencies between instructions, subsequently forming attention masks.
Our pre-trained model is general and can serve a wide range of applications.
The empirical results show that our technique substantially outperforms the state-of-the-art, as well as GraphCodeBERT-like pre-trained models, in three downstream tasks, including binary similarity, malware family classification, and type inference for binary executables.

\begin{acks}
We thank the anonymous reviewers for their valuable
comments and suggestions. This research was supported, in part by DARPA VSPELLS - HR001120S0058, IARPA TrojAI W911NF-19-S-0012, NSF 1901242 and 1910300, ONR N000141712045, N000141410468 and N000141712947. Any opinions, findings, and conclusions in this paper are those of the authors only and do not necessarily reflect the views of our sponsors.
\end{acks}

\bibliographystyle{ACM-Reference-Format}
\bibliography{reference}


\begin{thebibliography}{86}


\ifx \showCODEN    \undefined \def \showCODEN     #1{\unskip}     \fi
\ifx \showDOI      \undefined \def \showDOI       #1{#1}\fi
\ifx \showISBNx    \undefined \def \showISBNx     #1{\unskip}     \fi
\ifx \showISBNxiii \undefined \def \showISBNxiii  #1{\unskip}     \fi
\ifx \showISSN     \undefined \def \showISSN      #1{\unskip}     \fi
\ifx \showLCCN     \undefined \def \showLCCN      #1{\unskip}     \fi
\ifx \shownote     \undefined \def \shownote      #1{#1}          \fi
\ifx \showarticletitle \undefined \def \showarticletitle #1{#1}   \fi
\ifx \showURL      \undefined \def \showURL       {\relax}        \fi
\providecommand\bibfield[2]{#2}
\providecommand\bibinfo[2]{#2}
\providecommand\natexlab[1]{#1}
\providecommand\showeprint[2][]{arXiv:#2}

\bibitem[Achiam et~al\mbox{.}(2023)]%
        {achiam2023gpt}
\bibfield{author}{\bibinfo{person}{Josh Achiam}, \bibinfo{person}{Steven Adler}, \bibinfo{person}{Sandhini Agarwal}, \bibinfo{person}{Lama Ahmad}, \bibinfo{person}{Ilge Akkaya}, \bibinfo{person}{Florencia~Leoni Aleman}, \bibinfo{person}{Diogo Almeida}, \bibinfo{person}{Janko Altenschmidt}, \bibinfo{person}{Sam Altman}, \bibinfo{person}{Shyamal Anadkat}, {et~al\mbox{.}}} \bibinfo{year}{2023}\natexlab{}.
\newblock \showarticletitle{Gpt-4 technical report}.
\newblock \bibinfo{journal}{\emph{arXiv preprint arXiv:2303.08774}} (\bibinfo{year}{2023}).
\newblock


\bibitem[Ahmad et~al\mbox{.}(2021)]%
        {ahmad2021unified}
\bibfield{author}{\bibinfo{person}{Wasi Ahmad}, \bibinfo{person}{Saikat Chakraborty}, \bibinfo{person}{Baishakhi Ray}, {and} \bibinfo{person}{Kai-Wei Chang}.} \bibinfo{year}{2021}\natexlab{}.
\newblock \showarticletitle{Unified Pre-training for Program Understanding and Generation}. In \bibinfo{booktitle}{\emph{Proceedings of the 2021 Conference of the North American Chapter of the Association for Computational Linguistics: Human Language Technologies}}. \bibinfo{pages}{2655--2668}.
\newblock


\bibitem[Allamanis et~al\mbox{.}(2017)]%
        {allamanis2017learning}
\bibfield{author}{\bibinfo{person}{Miltiadis Allamanis}, \bibinfo{person}{Marc Brockschmidt}, {and} \bibinfo{person}{Mahmoud Khademi}.} \bibinfo{year}{2017}\natexlab{}.
\newblock \showarticletitle{Learning to represent programs with graphs}.
\newblock \bibinfo{journal}{\emph{arXiv preprint arXiv:1711.00740}} (\bibinfo{year}{2017}).
\newblock


\bibitem[Allen and Cocke(1976)]%
        {allen1976program}
\bibfield{author}{\bibinfo{person}{Frances~E. Allen} {and} \bibinfo{person}{John Cocke}.} \bibinfo{year}{1976}\natexlab{}.
\newblock \showarticletitle{A program data flow analysis procedure}.
\newblock \bibinfo{journal}{\emph{Commun. ACM}} \bibinfo{volume}{19}, \bibinfo{number}{3} (\bibinfo{year}{1976}), \bibinfo{pages}{137}.
\newblock


\bibitem[Alon and Yahav(2020)]%
        {alon2020bottleneck}
\bibfield{author}{\bibinfo{person}{Uri Alon} {and} \bibinfo{person}{Eran Yahav}.} \bibinfo{year}{2020}\natexlab{}.
\newblock \showarticletitle{On the bottleneck of graph neural networks and its practical implications}.
\newblock \bibinfo{journal}{\emph{arXiv preprint arXiv:2006.05205}} (\bibinfo{year}{2020}).
\newblock


\bibitem[Bieber et~al\mbox{.}(2020)]%
        {bieber2020learning}
\bibfield{author}{\bibinfo{person}{David Bieber}, \bibinfo{person}{Charles Sutton}, \bibinfo{person}{Hugo Larochelle}, {and} \bibinfo{person}{Daniel Tarlow}.} \bibinfo{year}{2020}\natexlab{}.
\newblock \showarticletitle{Learning to execute programs with instruction pointer attention graph neural networks}.
\newblock \bibinfo{journal}{\emph{Advances in Neural Information Processing Systems}}  \bibinfo{volume}{33} (\bibinfo{year}{2020}), \bibinfo{pages}{8626--8637}.
\newblock


\bibitem[Brown et~al\mbox{.}(2020)]%
        {brown2020language}
\bibfield{author}{\bibinfo{person}{Tom Brown}, \bibinfo{person}{Benjamin Mann}, \bibinfo{person}{Nick Ryder}, \bibinfo{person}{Melanie Subbiah}, \bibinfo{person}{Jared~D Kaplan}, \bibinfo{person}{Prafulla Dhariwal}, \bibinfo{person}{Arvind Neelakantan}, \bibinfo{person}{Pranav Shyam}, \bibinfo{person}{Girish Sastry}, \bibinfo{person}{Amanda Askell}, {et~al\mbox{.}}} \bibinfo{year}{2020}\natexlab{}.
\newblock \showarticletitle{Language models are few-shot learners}.
\newblock \bibinfo{journal}{\emph{Advances in neural information processing systems}}  \bibinfo{volume}{33} (\bibinfo{year}{2020}), \bibinfo{pages}{1877--1901}.
\newblock


\bibitem[Bui et~al\mbox{.}(2021a)]%
        {bui2021infercode}
\bibfield{author}{\bibinfo{person}{Nghi~DQ Bui}, \bibinfo{person}{Yijun Yu}, {and} \bibinfo{person}{Lingxiao Jiang}.} \bibinfo{year}{2021}\natexlab{a}.
\newblock \showarticletitle{Infercode: Self-supervised learning of code representations by predicting subtrees}. In \bibinfo{booktitle}{\emph{2021 IEEE/ACM 43rd International Conference on Software Engineering (ICSE)}}. IEEE, \bibinfo{pages}{1186--1197}.
\newblock


\bibitem[Bui et~al\mbox{.}(2021b)]%
        {bui2021treecaps}
\bibfield{author}{\bibinfo{person}{Nghi~DQ Bui}, \bibinfo{person}{Yijun Yu}, {and} \bibinfo{person}{Lingxiao Jiang}.} \bibinfo{year}{2021}\natexlab{b}.
\newblock \showarticletitle{Treecaps: Tree-based capsule networks for source code processing}. In \bibinfo{booktitle}{\emph{Proceedings of the AAAI Conference on Artificial Intelligence}}, Vol.~\bibinfo{volume}{35}. \bibinfo{pages}{30--38}.
\newblock


\bibitem[Cao et~al\mbox{.}(2023)]%
        {cao2023study}
\bibfield{author}{\bibinfo{person}{Jialun Cao}, \bibinfo{person}{Meiziniu Li}, \bibinfo{person}{Ming Wen}, {and} \bibinfo{person}{Shing-chi Cheung}.} \bibinfo{year}{2023}\natexlab{}.
\newblock \showarticletitle{A study on prompt design, advantages and limitations of chatgpt for deep learning program repair}.
\newblock \bibinfo{journal}{\emph{arXiv preprint arXiv:2304.08191}} (\bibinfo{year}{2023}).
\newblock


\bibitem[Cao et~al\mbox{.}(2022)]%
        {cao2022galois}
\bibfield{author}{\bibinfo{person}{Yushi Cao}, \bibinfo{person}{Zhiming Li}, \bibinfo{person}{Tianpei Yang}, \bibinfo{person}{Hao Zhang}, \bibinfo{person}{Yan Zheng}, \bibinfo{person}{Yi Li}, \bibinfo{person}{Jianye Hao}, {and} \bibinfo{person}{Yang Liu}.} \bibinfo{year}{2022}\natexlab{}.
\newblock \showarticletitle{GALOIS: boosting deep reinforcement learning via generalizable logic synthesis}.
\newblock \bibinfo{journal}{\emph{Advances in Neural Information Processing Systems}}  \bibinfo{volume}{35} (\bibinfo{year}{2022}), \bibinfo{pages}{19930--19943}.
\newblock


\bibitem[Chakraborty et~al\mbox{.}(2022)]%
        {chakraborty2022natgen}
\bibfield{author}{\bibinfo{person}{Saikat Chakraborty}, \bibinfo{person}{Toufique Ahmed}, \bibinfo{person}{Yangruibo Ding}, \bibinfo{person}{Premkumar~T Devanbu}, {and} \bibinfo{person}{Baishakhi Ray}.} \bibinfo{year}{2022}\natexlab{}.
\newblock \showarticletitle{Natgen: generative pre-training by “naturalizing” source code}. In \bibinfo{booktitle}{\emph{Proceedings of the 30th ACM Joint European Software Engineering Conference and Symposium on the Foundations of Software Engineering}}. \bibinfo{pages}{18--30}.
\newblock


\bibitem[Chen et~al\mbox{.}(2020)]%
        {chen2020measuring}
\bibfield{author}{\bibinfo{person}{Deli Chen}, \bibinfo{person}{Yankai Lin}, \bibinfo{person}{Wei Li}, \bibinfo{person}{Peng Li}, \bibinfo{person}{Jie Zhou}, {and} \bibinfo{person}{Xu Sun}.} \bibinfo{year}{2020}\natexlab{}.
\newblock \showarticletitle{Measuring and relieving the over-smoothing problem for graph neural networks from the topological view}. In \bibinfo{booktitle}{\emph{Proceedings of the AAAI conference on artificial intelligence}}, Vol.~\bibinfo{volume}{34}. \bibinfo{pages}{3438--3445}.
\newblock


\bibitem[Chen et~al\mbox{.}(2022a)]%
        {chen2022transferability}
\bibfield{author}{\bibinfo{person}{Fuxiang Chen}, \bibinfo{person}{Fatemeh~H Fard}, \bibinfo{person}{David Lo}, {and} \bibinfo{person}{Timofey Bryksin}.} \bibinfo{year}{2022}\natexlab{a}.
\newblock \showarticletitle{On the transferability of pre-trained language models for low-resource programming languages}. In \bibinfo{booktitle}{\emph{Proceedings of the 30th IEEE/ACM International Conference on Program Comprehension}}. \bibinfo{pages}{401--412}.
\newblock


\bibitem[Chen et~al\mbox{.}(2022b)]%
        {chen2022augmenting}
\bibfield{author}{\bibinfo{person}{Qibin Chen}, \bibinfo{person}{Jeremy Lacomis}, \bibinfo{person}{Edward~J Schwartz}, \bibinfo{person}{Claire Le~Goues}, \bibinfo{person}{Graham Neubig}, {and} \bibinfo{person}{Bogdan Vasilescu}.} \bibinfo{year}{2022}\natexlab{b}.
\newblock \showarticletitle{Augmenting decompiler output with learned variable names and types}. In \bibinfo{booktitle}{\emph{31st USENIX Security Symposium (USENIX Security 22)}}. \bibinfo{pages}{4327--4343}.
\newblock


\bibitem[Chen et~al\mbox{.}(2021)]%
        {chen2021plur}
\bibfield{author}{\bibinfo{person}{Zimin Chen}, \bibinfo{person}{Vincent~J Hellendoorn}, \bibinfo{person}{Pascal Lamblin}, \bibinfo{person}{Petros Maniatis}, \bibinfo{person}{Pierre-Antoine Manzagol}, \bibinfo{person}{Daniel Tarlow}, {and} \bibinfo{person}{Subhodeep Moitra}.} \bibinfo{year}{2021}\natexlab{}.
\newblock \showarticletitle{PLUR: A unifying, graph-based view of program learning, understanding, and repair}.
\newblock \bibinfo{journal}{\emph{Advances in Neural Information Processing Systems}}  \bibinfo{volume}{34} (\bibinfo{year}{2021}), \bibinfo{pages}{23089--23101}.
\newblock


\bibitem[Cormen et~al\mbox{.}(2022)]%
        {cormen2022introduction}
\bibfield{author}{\bibinfo{person}{Thomas~H Cormen}, \bibinfo{person}{Charles~E Leiserson}, \bibinfo{person}{Ronald~L Rivest}, {and} \bibinfo{person}{Clifford Stein}.} \bibinfo{year}{2022}\natexlab{}.
\newblock \bibinfo{booktitle}{\emph{Introduction to algorithms}}.
\newblock \bibinfo{publisher}{MIT press}.
\newblock


\bibitem[Devlin et~al\mbox{.}(2018)]%
        {devlin2018bert}
\bibfield{author}{\bibinfo{person}{Jacob Devlin}, \bibinfo{person}{Ming-Wei Chang}, \bibinfo{person}{Kenton Lee}, {and} \bibinfo{person}{Kristina Toutanova}.} \bibinfo{year}{2018}\natexlab{}.
\newblock \showarticletitle{Bert: Pre-training of deep bidirectional transformers for language understanding}.
\newblock \bibinfo{journal}{\emph{arXiv preprint arXiv:1810.04805}} (\bibinfo{year}{2018}).
\newblock


\bibitem[Ding et~al\mbox{.}(2023b)]%
        {ding2023traced}
\bibfield{author}{\bibinfo{person}{Yangruibo Ding}, \bibinfo{person}{Ben Steenhoek}, \bibinfo{person}{Kexin Pei}, \bibinfo{person}{Gail Kaiser}, \bibinfo{person}{Wei Le}, {and} \bibinfo{person}{Baishakhi Ray}.} \bibinfo{year}{2023}\natexlab{b}.
\newblock \showarticletitle{TRACED: Execution-aware Pre-training for Source Code}.
\newblock \bibinfo{journal}{\emph{arXiv preprint arXiv:2306.07487}} (\bibinfo{year}{2023}).
\newblock


\bibitem[Ding et~al\mbox{.}(2023a)]%
        {ding2023towards}
\bibfield{author}{\bibinfo{person}{Zishuo Ding}, \bibinfo{person}{Heng Li}, \bibinfo{person}{Weiyi Shang}, {and} \bibinfo{person}{Tse-Hsun Chen}.} \bibinfo{year}{2023}\natexlab{a}.
\newblock \showarticletitle{Towards learning generalizable code embeddings using task-agnostic graph convolutional networks}.
\newblock \bibinfo{journal}{\emph{ACM Transactions on Software Engineering and Methodology}} \bibinfo{volume}{32}, \bibinfo{number}{2} (\bibinfo{year}{2023}), \bibinfo{pages}{1--43}.
\newblock


\bibitem[Ding et~al\mbox{.}(2022)]%
        {ding2022can}
\bibfield{author}{\bibinfo{person}{Zishuo Ding}, \bibinfo{person}{Heng Li}, \bibinfo{person}{Weiyi Shang}, {and} \bibinfo{person}{Tse-Hsun~Peter Chen}.} \bibinfo{year}{2022}\natexlab{}.
\newblock \showarticletitle{Can pre-trained code embeddings improve model performance? Revisiting the use of code embeddings in software engineering tasks}.
\newblock \bibinfo{journal}{\emph{Empirical Software Engineering}} \bibinfo{volume}{27}, \bibinfo{number}{3} (\bibinfo{year}{2022}), \bibinfo{pages}{63}.
\newblock


\bibitem[Dong et~al\mbox{.}(2022)]%
        {dong2022fira}
\bibfield{author}{\bibinfo{person}{Jinhao Dong}, \bibinfo{person}{Yiling Lou}, \bibinfo{person}{Qihao Zhu}, \bibinfo{person}{Zeyu Sun}, \bibinfo{person}{Zhilin Li}, \bibinfo{person}{Wenjie Zhang}, {and} \bibinfo{person}{Dan Hao}.} \bibinfo{year}{2022}\natexlab{}.
\newblock \showarticletitle{FIRA: fine-grained graph-based code change representation for automated commit message generation}. In \bibinfo{booktitle}{\emph{Proceedings of the 44th International Conference on Software Engineering}}. \bibinfo{pages}{970--981}.
\newblock


\bibitem[Duan et~al\mbox{.}(2020)]%
        {duan2020deepbindiff}
\bibfield{author}{\bibinfo{person}{Yue Duan}, \bibinfo{person}{Xuezixiang Li}, \bibinfo{person}{Jinghan Wang}, {and} \bibinfo{person}{Heng Yin}.} \bibinfo{year}{2020}\natexlab{}.
\newblock \showarticletitle{Deepbindiff: Learning program-wide code representations for binary diffing}. In \bibinfo{booktitle}{\emph{Network and distributed system security symposium}}.
\newblock


\bibitem[Fan et~al\mbox{.}(2023)]%
        {fan2023automated}
\bibfield{author}{\bibinfo{person}{Zhiyu Fan}, \bibinfo{person}{Xiang Gao}, \bibinfo{person}{Martin Mirchev}, \bibinfo{person}{Abhik Roychoudhury}, {and} \bibinfo{person}{Shin~Hwei Tan}.} \bibinfo{year}{2023}\natexlab{}.
\newblock \showarticletitle{Automated repair of programs from large language models}. In \bibinfo{booktitle}{\emph{2023 IEEE/ACM 45th International Conference on Software Engineering (ICSE)}}. IEEE, \bibinfo{pages}{1469--1481}.
\newblock


\bibitem[Fang et~al\mbox{.}(2023b)]%
        {fang2023isyn}
\bibfield{author}{\bibinfo{person}{Pengcheng Fang}, \bibinfo{person}{Zhenhua Zou}, \bibinfo{person}{Xusheng Xiao}, {and} \bibinfo{person}{Zhuotao Liu}.} \bibinfo{year}{2023}\natexlab{b}.
\newblock \showarticletitle{iSyn: Semi-automated Smart Contract Synthesis from Legal Financial Agreements}. In \bibinfo{booktitle}{\emph{Proceedings of the 32nd ACM SIGSOFT International Symposium on Software Testing and Analysis}}. \bibinfo{pages}{727--739}.
\newblock


\bibitem[Fang et~al\mbox{.}(2023a)]%
        {fang2023representthemall}
\bibfield{author}{\bibinfo{person}{Sen Fang}, \bibinfo{person}{Tao Zhang}, \bibinfo{person}{Youshuai Tan}, \bibinfo{person}{He Jiang}, \bibinfo{person}{Xin Xia}, {and} \bibinfo{person}{Xiaobing Sun}.} \bibinfo{year}{2023}\natexlab{a}.
\newblock \showarticletitle{RepresentThemAll: A Universal Learning Representation of Bug Reports}. In \bibinfo{booktitle}{\emph{2023 IEEE/ACM 45th International Conference on Software Engineering (ICSE)}}. IEEE, \bibinfo{pages}{602--614}.
\newblock


\bibitem[Fawaz et~al\mbox{.}(2015)]%
        {fawaz2015anatomization}
\bibfield{author}{\bibinfo{person}{Kassem Fawaz}, \bibinfo{person}{Huan Feng}, {and} \bibinfo{person}{Kang~G Shin}.} \bibinfo{year}{2015}\natexlab{}.
\newblock \showarticletitle{Anatomization and protection of mobile apps’ location privacy threats}. In \bibinfo{booktitle}{\emph{24th USENIX Security Symposium (USENIX Security 15)}}. \bibinfo{pages}{753--768}.
\newblock


\bibitem[Feng et~al\mbox{.}(2020)]%
        {feng2020codebert}
\bibfield{author}{\bibinfo{person}{Zhangyin Feng}, \bibinfo{person}{Daya Guo}, \bibinfo{person}{Duyu Tang}, \bibinfo{person}{Nan Duan}, \bibinfo{person}{Xiaocheng Feng}, \bibinfo{person}{Ming Gong}, \bibinfo{person}{Linjun Shou}, \bibinfo{person}{Bing Qin}, \bibinfo{person}{Ting Liu}, \bibinfo{person}{Daxin Jiang}, {et~al\mbox{.}}} \bibinfo{year}{2020}\natexlab{}.
\newblock \showarticletitle{CodeBERT: A Pre-Trained Model for Programming and Natural Languages}. In \bibinfo{booktitle}{\emph{Findings of the Association for Computational Linguistics: EMNLP 2020}}. \bibinfo{pages}{1536--1547}.
\newblock


\bibitem[Ferrante et~al\mbox{.}(1987)]%
        {ferrante1987program}
\bibfield{author}{\bibinfo{person}{Jeanne Ferrante}, \bibinfo{person}{Karl~J Ottenstein}, {and} \bibinfo{person}{Joe~D Warren}.} \bibinfo{year}{1987}\natexlab{}.
\newblock \showarticletitle{The program dependence graph and its use in optimization}.
\newblock \bibinfo{journal}{\emph{ACM Transactions on Programming Languages and Systems (TOPLAS)}} \bibinfo{volume}{9}, \bibinfo{number}{3} (\bibinfo{year}{1987}), \bibinfo{pages}{319--349}.
\newblock


\bibitem[Gao et~al\mbox{.}({[n.\,d.]})]%
        {gao2024sigmadiff}
\bibfield{author}{\bibinfo{person}{Lian Gao}, \bibinfo{person}{Yu Qu}, \bibinfo{person}{Sheng Yu}, \bibinfo{person}{Yue Duan}, {and} \bibinfo{person}{Heng Yin}.} \bibinfo{year}{[n.\,d.]}\natexlab{}.
\newblock \showarticletitle{SIGMADIFF: Semantics-Aware Deep Graph Matching for Pseudocode Diffing}. In \bibinfo{booktitle}{\emph{Network and Distributed System Security (NDSS) Symposium 2024}}.
\newblock


\bibitem[Gao et~al\mbox{.}(2023)]%
        {gao2023two}
\bibfield{author}{\bibinfo{person}{Shuzheng Gao}, \bibinfo{person}{Cuiyun Gao}, \bibinfo{person}{Chaozheng Wang}, \bibinfo{person}{Jun Sun}, \bibinfo{person}{David Lo}, {and} \bibinfo{person}{Yue Yu}.} \bibinfo{year}{2023}\natexlab{}.
\newblock \showarticletitle{Two sides of the same coin: Exploiting the impact of identifiers in neural code comprehension}. In \bibinfo{booktitle}{\emph{2023 IEEE/ACM 45th International Conference on Software Engineering (ICSE)}}. IEEE, \bibinfo{pages}{1933--1945}.
\newblock


\bibitem[Gray et~al\mbox{.}(2021)]%
        {gray2021identifying}
\bibfield{author}{\bibinfo{person}{Jason Gray}, \bibinfo{person}{Daniele Sgandurra}, {and} \bibinfo{person}{Lorenzo Cavallaro}.} \bibinfo{year}{2021}\natexlab{}.
\newblock \showarticletitle{Identifying authorship style in malicious binaries: techniques, challenges \& datasets}.
\newblock \bibinfo{journal}{\emph{arXiv preprint arXiv:2101.06124}} (\bibinfo{year}{2021}).
\newblock


\bibitem[Guo et~al\mbox{.}(2022)]%
        {guo2022unixcoder}
\bibfield{author}{\bibinfo{person}{Daya Guo}, \bibinfo{person}{Shuai Lu}, \bibinfo{person}{Nan Duan}, \bibinfo{person}{Yanlin Wang}, \bibinfo{person}{Ming Zhou}, {and} \bibinfo{person}{Jian Yin}.} \bibinfo{year}{2022}\natexlab{}.
\newblock \showarticletitle{UniXcoder: Unified Cross-Modal Pre-training for Code Representation}. In \bibinfo{booktitle}{\emph{Proceedings of the 60th Annual Meeting of the Association for Computational Linguistics (Volume 1: Long Papers)}}. \bibinfo{pages}{7212--7225}.
\newblock


\bibitem[Guo et~al\mbox{.}(2020b)]%
        {guo2020graphcodebert}
\bibfield{author}{\bibinfo{person}{Daya Guo}, \bibinfo{person}{Shuo Ren}, \bibinfo{person}{Shuai Lu}, \bibinfo{person}{Zhangyin Feng}, \bibinfo{person}{Duyu Tang}, \bibinfo{person}{Shujie Liu}, \bibinfo{person}{Long Zhou}, \bibinfo{person}{Nan Duan}, \bibinfo{person}{Alexey Svyatkovskiy}, \bibinfo{person}{Shengyu Fu}, {et~al\mbox{.}}} \bibinfo{year}{2020}\natexlab{b}.
\newblock \showarticletitle{Graphcodebert: Pre-training code representations with data flow}.
\newblock \bibinfo{journal}{\emph{arXiv preprint arXiv:2009.08366}} (\bibinfo{year}{2020}).
\newblock


\bibitem[Guo et~al\mbox{.}(2023)]%
        {guo2023study}
\bibfield{author}{\bibinfo{person}{Yuxiang Guo}, \bibinfo{person}{Xiaopeng Gao}, \bibinfo{person}{Zhenyu Zhang}, \bibinfo{person}{WK Chan}, {and} \bibinfo{person}{Bo Jiang}.} \bibinfo{year}{2023}\natexlab{}.
\newblock \showarticletitle{A study on the impact of pre-trained model on Just-In-Time defect prediction}.
\newblock \bibinfo{journal}{\emph{arXiv preprint arXiv:2309.02317}} (\bibinfo{year}{2023}).
\newblock


\bibitem[Guo et~al\mbox{.}(2020a)]%
        {guo2020graphspy}
\bibfield{author}{\bibinfo{person}{Yixin Guo}, \bibinfo{person}{Pengcheng Li}, \bibinfo{person}{Yingwei Luo}, \bibinfo{person}{Xiaolin Wang}, {and} \bibinfo{person}{Zhenlin Wang}.} \bibinfo{year}{2020}\natexlab{a}.
\newblock \showarticletitle{GRAPHSPY: Fused Program Semantic-Level Embedding via Graph Neural Networks for Dead Store Detection}.
\newblock \bibinfo{journal}{\emph{arXiv preprint arXiv:2011.09501}} (\bibinfo{year}{2020}).
\newblock


\bibitem[IDA Pro(2023)]%
        {idapro}
IDA Pro \bibinfo{year}{2023}\natexlab{}.
\newblock \bibinfo{booktitle}{\emph{A powerful disassembler and a versatile debugger}}.
\newblock
\urldef\tempurl%
\url{https://hex-rays.com/ida-pro/}
\showURL{%
\tempurl}


\bibitem[Jin et~al\mbox{.}(2022)]%
        {jin2022symlm}
\bibfield{author}{\bibinfo{person}{Xin Jin}, \bibinfo{person}{Kexin Pei}, \bibinfo{person}{Jun~Yeon Won}, {and} \bibinfo{person}{Zhiqiang Lin}.} \bibinfo{year}{2022}\natexlab{}.
\newblock \showarticletitle{Symlm: Predicting function names in stripped binaries via context-sensitive execution-aware code embeddings}. In \bibinfo{booktitle}{\emph{Proceedings of the 2022 ACM SIGSAC Conference on Computer and Communications Security}}. \bibinfo{pages}{1631--1645}.
\newblock


\bibitem[Kanade et~al\mbox{.}(2020)]%
        {kanade2020learning}
\bibfield{author}{\bibinfo{person}{Aditya Kanade}, \bibinfo{person}{Petros Maniatis}, \bibinfo{person}{Gogul Balakrishnan}, {and} \bibinfo{person}{Kensen Shi}.} \bibinfo{year}{2020}\natexlab{}.
\newblock \showarticletitle{Learning and evaluating contextual embedding of source code}. In \bibinfo{booktitle}{\emph{International conference on machine learning}}. PMLR, \bibinfo{pages}{5110--5121}.
\newblock


\bibitem[Kim et~al\mbox{.}(2022)]%
        {kim2022improving}
\bibfield{author}{\bibinfo{person}{Geunwoo Kim}, \bibinfo{person}{Sanghyun Hong}, \bibinfo{person}{Michael Franz}, {and} \bibinfo{person}{Dokyung Song}.} \bibinfo{year}{2022}\natexlab{}.
\newblock \showarticletitle{Improving cross-platform binary analysis using representation learning via graph alignment}. In \bibinfo{booktitle}{\emph{Proceedings of the 31st ACM SIGSOFT International Symposium on Software Testing and Analysis}}. \bibinfo{pages}{151--163}.
\newblock


\bibitem[Li et~al\mbox{.}(2018)]%
        {li2018deeper}
\bibfield{author}{\bibinfo{person}{Qimai Li}, \bibinfo{person}{Zhichao Han}, {and} \bibinfo{person}{Xiao-Ming Wu}.} \bibinfo{year}{2018}\natexlab{}.
\newblock \showarticletitle{Deeper insights into graph convolutional networks for semi-supervised learning}. In \bibinfo{booktitle}{\emph{Proceedings of the AAAI conference on artificial intelligence}}, Vol.~\bibinfo{volume}{32}.
\newblock


\bibitem[Li et~al\mbox{.}(2022)]%
        {li2022soft}
\bibfield{author}{\bibinfo{person}{Xiaonan Li}, \bibinfo{person}{Daya Guo}, \bibinfo{person}{Yeyun Gong}, \bibinfo{person}{Yun Lin}, \bibinfo{person}{Yelong Shen}, \bibinfo{person}{Xipeng Qiu}, \bibinfo{person}{Daxin Jiang}, \bibinfo{person}{Weizhu Chen}, {and} \bibinfo{person}{Nan Duan}.} \bibinfo{year}{2022}\natexlab{}.
\newblock \showarticletitle{Soft-Labeled Contrastive Pre-training for Function-level Code Representation}.
\newblock \bibinfo{journal}{\emph{arXiv preprint arXiv:2210.09597}} (\bibinfo{year}{2022}).
\newblock


\bibitem[Li et~al\mbox{.}(2019)]%
        {li2019graph}
\bibfield{author}{\bibinfo{person}{Yujia Li}, \bibinfo{person}{Chenjie Gu}, \bibinfo{person}{Thomas Dullien}, \bibinfo{person}{Oriol Vinyals}, {and} \bibinfo{person}{Pushmeet Kohli}.} \bibinfo{year}{2019}\natexlab{}.
\newblock \showarticletitle{Graph matching networks for learning the similarity of graph structured objects}. In \bibinfo{booktitle}{\emph{International conference on machine learning}}. PMLR, \bibinfo{pages}{3835--3845}.
\newblock


\bibitem[Lin et~al\mbox{.}(2023)]%
        {lin2023cct5}
\bibfield{author}{\bibinfo{person}{Bo Lin}, \bibinfo{person}{Shangwen Wang}, \bibinfo{person}{Zhongxin Liu}, \bibinfo{person}{Yepang Liu}, \bibinfo{person}{Xin Xia}, {and} \bibinfo{person}{Xiaoguang Mao}.} \bibinfo{year}{2023}\natexlab{}.
\newblock \showarticletitle{CCT5: A Code-Change-Oriented Pre-Trained Model}.
\newblock \bibinfo{journal}{\emph{arXiv preprint arXiv:2305.10785}} (\bibinfo{year}{2023}).
\newblock


\bibitem[Lin et~al\mbox{.}(2010)]%
        {lin2010automatic}
\bibfield{author}{\bibinfo{person}{Zhiqiang Lin}, \bibinfo{person}{Xiangyu Zhang}, {and} \bibinfo{person}{Dongyan Xu}.} \bibinfo{year}{2010}\natexlab{}.
\newblock \showarticletitle{Automatic reverse engineering of data structures from binary execution}. In \bibinfo{booktitle}{\emph{Proceedings of the 11th Annual Information Security Symposium}}. \bibinfo{pages}{1--1}.
\newblock


\bibitem[Liu et~al\mbox{.}(2019)]%
        {liu2019roberta}
\bibfield{author}{\bibinfo{person}{Yinhan Liu}, \bibinfo{person}{Myle Ott}, \bibinfo{person}{Naman Goyal}, \bibinfo{person}{Jingfei Du}, \bibinfo{person}{Mandar Joshi}, \bibinfo{person}{Danqi Chen}, \bibinfo{person}{Omer Levy}, \bibinfo{person}{Mike Lewis}, \bibinfo{person}{Luke Zettlemoyer}, {and} \bibinfo{person}{Veselin Stoyanov}.} \bibinfo{year}{2019}\natexlab{}.
\newblock \showarticletitle{Roberta: A robustly optimized bert pretraining approach}.
\newblock \bibinfo{journal}{\emph{arXiv preprint arXiv:1907.11692}} (\bibinfo{year}{2019}).
\newblock


\bibitem[Loshchilov and Hutter(2017)]%
        {loshchilov2017decoupled}
\bibfield{author}{\bibinfo{person}{Ilya Loshchilov} {and} \bibinfo{person}{Frank Hutter}.} \bibinfo{year}{2017}\natexlab{}.
\newblock \showarticletitle{Decoupled weight decay regularization}.
\newblock \bibinfo{journal}{\emph{arXiv preprint arXiv:1711.05101}} (\bibinfo{year}{2017}).
\newblock


\bibitem[Lu et~al\mbox{.}(2021)]%
        {lu2021codexglue}
\bibfield{author}{\bibinfo{person}{Shuai Lu}, \bibinfo{person}{Daya Guo}, \bibinfo{person}{Shuo Ren}, \bibinfo{person}{Junjie Huang}, \bibinfo{person}{Alexey Svyatkovskiy}, \bibinfo{person}{Ambrosio Blanco}, \bibinfo{person}{Colin Clement}, \bibinfo{person}{Dawn Drain}, \bibinfo{person}{Daxin Jiang}, \bibinfo{person}{Duyu Tang}, {et~al\mbox{.}}} \bibinfo{year}{2021}\natexlab{}.
\newblock \showarticletitle{Codexglue: A machine learning benchmark dataset for code understanding and generation}.
\newblock \bibinfo{journal}{\emph{arXiv preprint arXiv:2102.04664}} (\bibinfo{year}{2021}).
\newblock


\bibitem[Luo et~al\mbox{.}(2017)]%
        {luo2017semantics}
\bibfield{author}{\bibinfo{person}{Lannan Luo}, \bibinfo{person}{Jiang Ming}, \bibinfo{person}{Dinghao Wu}, \bibinfo{person}{Peng Liu}, {and} \bibinfo{person}{Sencun Zhu}.} \bibinfo{year}{2017}\natexlab{}.
\newblock \showarticletitle{Semantics-Based Obfuscation-Resilient Binary Code Similarity Comparison with Applications to Software and Algorithm Plagiarism Detection}.
\newblock \bibinfo{journal}{\emph{IEEE Transactions on Software Engineering}} \bibinfo{volume}{43}, \bibinfo{number}{12} (\bibinfo{year}{2017}), \bibinfo{pages}{1157--1177}.
\newblock
\urldef\tempurl%
\url{https://doi.org/10.1109/TSE.2017.2655046}
\showDOI{\tempurl}


\bibitem[Luong et~al\mbox{.}(2015)]%
        {luong2015effective}
\bibfield{author}{\bibinfo{person}{Minh-Thang Luong}, \bibinfo{person}{Hieu Pham}, {and} \bibinfo{person}{Christopher~D Manning}.} \bibinfo{year}{2015}\natexlab{}.
\newblock \showarticletitle{Effective Approaches to Attention-based Neural Machine Translation}. In \bibinfo{booktitle}{\emph{Proceedings of the 2015 Conference on Empirical Methods in Natural Language Processing}}. \bibinfo{pages}{1412--1421}.
\newblock


\bibitem[Ma et~al\mbox{.}(2022)]%
        {ma2022graphcode2vec}
\bibfield{author}{\bibinfo{person}{Wei Ma}, \bibinfo{person}{Mengjie Zhao}, \bibinfo{person}{Ezekiel Soremekun}, \bibinfo{person}{Qiang Hu}, \bibinfo{person}{Jie~M Zhang}, \bibinfo{person}{Mike Papadakis}, \bibinfo{person}{Maxime Cordy}, \bibinfo{person}{Xiaofei Xie}, {and} \bibinfo{person}{Yves~Le Traon}.} \bibinfo{year}{2022}\natexlab{}.
\newblock \showarticletitle{Graphcode2vec: Generic code embedding via lexical and program dependence analyses}. In \bibinfo{booktitle}{\emph{Proceedings of the 19th International Conference on Mining Software Repositories}}. \bibinfo{pages}{524--536}.
\newblock


\bibitem[Ma et~al\mbox{.}(2021)]%
        {ma2021comprehensive}
\bibfield{author}{\bibinfo{person}{Yixuan Ma}, \bibinfo{person}{Shuang Liu}, \bibinfo{person}{Jiajun Jiang}, \bibinfo{person}{Guanhong Chen}, {and} \bibinfo{person}{Keqiu Li}.} \bibinfo{year}{2021}\natexlab{}.
\newblock \showarticletitle{A comprehensive study on learning-based PE malware family classification methods}. In \bibinfo{booktitle}{\emph{Proceedings of the 29th ACM Joint Meeting on European Software Engineering Conference and Symposium on the Foundations of Software Engineering}}. \bibinfo{pages}{1314--1325}.
\newblock


\bibitem[Marcelli et~al\mbox{.}(2022)]%
        {marcelli2022machine}
\bibfield{author}{\bibinfo{person}{Andrea Marcelli}, \bibinfo{person}{Mariano Graziano}, \bibinfo{person}{Xabier Ugarte-Pedrero}, \bibinfo{person}{Yanick Fratantonio}, \bibinfo{person}{Mohamad Mansouri}, {and} \bibinfo{person}{Davide Balzarotti}.} \bibinfo{year}{2022}\natexlab{}.
\newblock \showarticletitle{How machine learning is solving the binary function similarity problem}. In \bibinfo{booktitle}{\emph{31st USENIX Security Symposium (USENIX Security 22)}}. \bibinfo{pages}{2099--2116}.
\newblock


\bibitem[Napier et~al\mbox{.}(2023)]%
        {napier2023empirical}
\bibfield{author}{\bibinfo{person}{Kollin Napier}, \bibinfo{person}{Tanmay Bhowmik}, {and} \bibinfo{person}{Shaowei Wang}.} \bibinfo{year}{2023}\natexlab{}.
\newblock \showarticletitle{An empirical study of text-based machine learning models for vulnerability detection}.
\newblock \bibinfo{journal}{\emph{Empirical Software Engineering}} \bibinfo{volume}{28}, \bibinfo{number}{2} (\bibinfo{year}{2023}), \bibinfo{pages}{38}.
\newblock


\bibitem[Nataraj et~al\mbox{.}(2011)]%
        {nataraj2011malware}
\bibfield{author}{\bibinfo{person}{L. Nataraj}, \bibinfo{person}{S. Karthikeyan}, \bibinfo{person}{G. Jacob}, {and} \bibinfo{person}{B.~S. Manjunath}.} \bibinfo{year}{2011}\natexlab{}.
\newblock \showarticletitle{Malware Images: Visualization and Automatic Classification}. In \bibinfo{booktitle}{\emph{Proceedings of the 8th International Symposium on Visualization for Cyber Security}} (Pittsburgh, Pennsylvania, USA) \emph{(\bibinfo{series}{VizSec '11})}. \bibinfo{publisher}{Association for Computing Machinery}, \bibinfo{address}{New York, NY, USA}, Article \bibinfo{articleno}{4}, \bibinfo{numpages}{7}~pages.
\newblock
\showISBNx{9781450306799}
\urldef\tempurl%
\url{https://doi.org/10.1145/2016904.2016908}
\showDOI{\tempurl}


\bibitem[Pei et~al\mbox{.}(2021)]%
        {pei2021stateformer}
\bibfield{author}{\bibinfo{person}{Kexin Pei}, \bibinfo{person}{Jonas Guan}, \bibinfo{person}{Matthew Broughton}, \bibinfo{person}{Zhongtian Chen}, \bibinfo{person}{Songchen Yao}, \bibinfo{person}{David Williams-King}, \bibinfo{person}{Vikas Ummadisetty}, \bibinfo{person}{Junfeng Yang}, \bibinfo{person}{Baishakhi Ray}, {and} \bibinfo{person}{Suman Jana}.} \bibinfo{year}{2021}\natexlab{}.
\newblock \showarticletitle{StateFormer: Fine-grained type recovery from binaries using generative state modeling}. In \bibinfo{booktitle}{\emph{Proceedings of the 29th ACM Joint Meeting on European Software Engineering Conference and Symposium on the Foundations of Software Engineering}}. \bibinfo{pages}{690--702}.
\newblock


\bibitem[Pei et~al\mbox{.}(2020)]%
        {pei2020trex}
\bibfield{author}{\bibinfo{person}{Kexin Pei}, \bibinfo{person}{Zhou Xuan}, \bibinfo{person}{Junfeng Yang}, \bibinfo{person}{Suman Jana}, {and} \bibinfo{person}{Baishakhi Ray}.} \bibinfo{year}{2020}\natexlab{}.
\newblock \showarticletitle{Trex: Learning execution semantics from micro-traces for binary similarity}.
\newblock \bibinfo{journal}{\emph{arXiv preprint arXiv:2012.08680}} (\bibinfo{year}{2020}).
\newblock


\bibitem[Peng et~al\mbox{.}(2021)]%
        {peng2021could}
\bibfield{author}{\bibinfo{person}{Dinglan Peng}, \bibinfo{person}{Shuxin Zheng}, \bibinfo{person}{Yatao Li}, \bibinfo{person}{Guolin Ke}, \bibinfo{person}{Di He}, {and} \bibinfo{person}{Tie-Yan Liu}.} \bibinfo{year}{2021}\natexlab{}.
\newblock \showarticletitle{How could neural networks understand programs?}. In \bibinfo{booktitle}{\emph{International Conference on Machine Learning}}. PMLR, \bibinfo{pages}{8476--8486}.
\newblock


\bibitem[Peng et~al\mbox{.}(2023)]%
        {peng2023domain}
\bibfield{author}{\bibinfo{person}{Yun Peng}, \bibinfo{person}{Shuzheng Gao}, \bibinfo{person}{Cuiyun Gao}, \bibinfo{person}{Yintong Huo}, {and} \bibinfo{person}{Michael~R Lyu}.} \bibinfo{year}{2023}\natexlab{}.
\newblock \showarticletitle{Domain Knowledge Matters: Improving Prompts with Fix Templates for Repairing Python Type Errors}.
\newblock \bibinfo{journal}{\emph{arXiv preprint arXiv:2306.01394}} (\bibinfo{year}{2023}).
\newblock


\bibitem[Pierce(2002)]%
        {pierce2002types}
\bibfield{author}{\bibinfo{person}{Benjamin~C Pierce}.} \bibinfo{year}{2002}\natexlab{}.
\newblock \bibinfo{booktitle}{\emph{Types and programming languages}}.
\newblock \bibinfo{publisher}{MIT press}.
\newblock


\bibitem[Radford et~al\mbox{.}(2018)]%
        {radford2018improving}
\bibfield{author}{\bibinfo{person}{Alec Radford}, \bibinfo{person}{Karthik Narasimhan}, \bibinfo{person}{Tim Salimans}, \bibinfo{person}{Ilya Sutskever}, {et~al\mbox{.}}} \bibinfo{year}{2018}\natexlab{}.
\newblock \showarticletitle{Improving language understanding by generative pre-training}.
\newblock  (\bibinfo{year}{2018}).
\newblock


\bibitem[Raffel et~al\mbox{.}(2020)]%
        {raffel2020exploring}
\bibfield{author}{\bibinfo{person}{Colin Raffel}, \bibinfo{person}{Noam Shazeer}, \bibinfo{person}{Adam Roberts}, \bibinfo{person}{Katherine Lee}, \bibinfo{person}{Sharan Narang}, \bibinfo{person}{Michael Matena}, \bibinfo{person}{Yanqi Zhou}, \bibinfo{person}{Wei Li}, {and} \bibinfo{person}{Peter~J Liu}.} \bibinfo{year}{2020}\natexlab{}.
\newblock \showarticletitle{Exploring the limits of transfer learning with a unified text-to-text transformer}.
\newblock \bibinfo{journal}{\emph{The Journal of Machine Learning Research}} \bibinfo{volume}{21}, \bibinfo{number}{1} (\bibinfo{year}{2020}), \bibinfo{pages}{5485--5551}.
\newblock


\bibitem[Rozi{\`e}re et~al\mbox{.}(2023)]%
        {roziere2023code}
\bibfield{author}{\bibinfo{person}{Baptiste Rozi{\`e}re}, \bibinfo{person}{Jonas Gehring}, \bibinfo{person}{Fabian Gloeckle}, \bibinfo{person}{Sten Sootla}, \bibinfo{person}{Itai Gat}, \bibinfo{person}{Xiaoqing~Ellen Tan}, \bibinfo{person}{Yossi Adi}, \bibinfo{person}{Jingyu Liu}, \bibinfo{person}{Tal Remez}, \bibinfo{person}{J{\'e}r{\'e}my Rapin}, {et~al\mbox{.}}} \bibinfo{year}{2023}\natexlab{}.
\newblock \showarticletitle{Code llama: Open foundation models for code}.
\newblock \bibinfo{journal}{\emph{arXiv preprint arXiv:2308.12950}} (\bibinfo{year}{2023}).
\newblock


\bibitem[Saberi and Fard(2023)]%
        {saberi2023model}
\bibfield{author}{\bibinfo{person}{Iman Saberi} {and} \bibinfo{person}{Fatemeh~H Fard}.} \bibinfo{year}{2023}\natexlab{}.
\newblock \showarticletitle{Model-Agnostic Syntactical Information for Pre-Trained Programming Language Models}.
\newblock \bibinfo{journal}{\emph{arXiv preprint arXiv:2303.06233}} (\bibinfo{year}{2023}).
\newblock


\bibitem[Sattler et~al\mbox{.}(2023)]%
        {sattler2023seal}
\bibfield{author}{\bibinfo{person}{Florian Sattler}, \bibinfo{person}{Sebastian B{\"o}hm}, \bibinfo{person}{Philipp~Dominik Schubert}, \bibinfo{person}{Norbert Siegmund}, {and} \bibinfo{person}{Sven Apel}.} \bibinfo{year}{2023}\natexlab{}.
\newblock \showarticletitle{SEAL: Integrating Program Analysis and Repository Mining}.
\newblock \bibinfo{journal}{\emph{ACM Transactions on Software Engineering and Methodology}} (\bibinfo{year}{2023}).
\newblock


\bibitem[Shariffdeen et~al\mbox{.}(2021)]%
        {shari2021automated}
\bibfield{author}{\bibinfo{person}{Ridwan~Salihin Shariffdeen}, \bibinfo{person}{Shin~Hwei Tan}, \bibinfo{person}{Mingyuan Gao}, {and} \bibinfo{person}{Abhik Roychoudhury}.} \bibinfo{year}{2021}\natexlab{}.
\newblock \showarticletitle{Automated Patch Transplantation}.
\newblock \bibinfo{journal}{\emph{ACM Trans. Softw. Eng. Methodol.}} \bibinfo{volume}{30}, \bibinfo{number}{1}, Article \bibinfo{articleno}{6} (\bibinfo{date}{dec} \bibinfo{year}{2021}), \bibinfo{numpages}{36}~pages.
\newblock
\showISSN{1049-331X}
\urldef\tempurl%
\url{https://doi.org/10.1145/3412376}
\showDOI{\tempurl}


\bibitem[Shi et~al\mbox{.}(2023)]%
        {shi2023towards}
\bibfield{author}{\bibinfo{person}{Ensheng Shi}, \bibinfo{person}{Yanlin Wang}, \bibinfo{person}{Hongyu Zhang}, \bibinfo{person}{Lun Du}, \bibinfo{person}{Shi Han}, \bibinfo{person}{Dongmei Zhang}, {and} \bibinfo{person}{Hongbin Sun}.} \bibinfo{year}{2023}\natexlab{}.
\newblock \showarticletitle{Towards Efficient Fine-tuning of Pre-trained Code Models: An Experimental Study and Beyond}.
\newblock \bibinfo{journal}{\emph{arXiv preprint arXiv:2304.05216}} (\bibinfo{year}{2023}).
\newblock


\bibitem[Shi et~al\mbox{.}(2022)]%
        {shi2022better}
\bibfield{author}{\bibinfo{person}{Yucen Shi}, \bibinfo{person}{Ying Yin}, \bibinfo{person}{Zhengkui Wang}, \bibinfo{person}{David Lo}, \bibinfo{person}{Tao Zhang}, \bibinfo{person}{Xin Xia}, \bibinfo{person}{Yuhai Zhao}, {and} \bibinfo{person}{Bowen Xu}.} \bibinfo{year}{2022}\natexlab{}.
\newblock \showarticletitle{How to better utilize code graphs in semantic code search?}. In \bibinfo{booktitle}{\emph{Proceedings of the 30th ACM Joint European Software Engineering Conference and Symposium on the Foundations of Software Engineering}}. \bibinfo{pages}{722--733}.
\newblock


\bibitem[Steenhoek et~al\mbox{.}(2024)]%
        {steenhoek2024dataflow}
\bibfield{author}{\bibinfo{person}{Benjamin Steenhoek}, \bibinfo{person}{Hongyang Gao}, {and} \bibinfo{person}{Wei Le}.} \bibinfo{year}{2024}\natexlab{}.
\newblock \showarticletitle{Dataflow Analysis-Inspired Deep Learning for Efficient Vulnerability Detection}. In \bibinfo{booktitle}{\emph{Proceedings of the 46th IEEE/ACM International Conference on Software Engineering}}. \bibinfo{pages}{1--13}.
\newblock


\bibitem[Sun et~al\mbox{.}(2020)]%
        {sun2020treegen}
\bibfield{author}{\bibinfo{person}{Zeyu Sun}, \bibinfo{person}{Qihao Zhu}, \bibinfo{person}{Yingfei Xiong}, \bibinfo{person}{Yican Sun}, \bibinfo{person}{Lili Mou}, {and} \bibinfo{person}{Lu Zhang}.} \bibinfo{year}{2020}\natexlab{}.
\newblock \showarticletitle{Treegen: A tree-based transformer architecture for code generation}. In \bibinfo{booktitle}{\emph{Proceedings of the AAAI Conference on Artificial Intelligence}}, Vol.~\bibinfo{volume}{34}. \bibinfo{pages}{8984--8991}.
\newblock


\bibitem[Tsoumakas et~al\mbox{.}(2010)]%
        {tsoumakas2010mining}
\bibfield{author}{\bibinfo{person}{Grigorios Tsoumakas}, \bibinfo{person}{Ioannis Katakis}, {and} \bibinfo{person}{Ioannis Vlahavas}.} \bibinfo{year}{2010}\natexlab{}.
\newblock \showarticletitle{Mining multi-label data}.
\newblock \bibinfo{journal}{\emph{Data mining and knowledge discovery handbook}} (\bibinfo{year}{2010}), \bibinfo{pages}{667--685}.
\newblock


\bibitem[Vaswani et~al\mbox{.}(2017)]%
        {vaswani2017attention}
\bibfield{author}{\bibinfo{person}{Ashish Vaswani}, \bibinfo{person}{Noam Shazeer}, \bibinfo{person}{Niki Parmar}, \bibinfo{person}{Jakob Uszkoreit}, \bibinfo{person}{Llion Jones}, \bibinfo{person}{Aidan~N Gomez}, \bibinfo{person}{{\L}ukasz Kaiser}, {and} \bibinfo{person}{Illia Polosukhin}.} \bibinfo{year}{2017}\natexlab{}.
\newblock \showarticletitle{Attention is all you need}.
\newblock \bibinfo{journal}{\emph{Advances in neural information processing systems}}  \bibinfo{volume}{30} (\bibinfo{year}{2017}).
\newblock


\bibitem[Wang et~al\mbox{.}(2022a)]%
        {wang2022bridging}
\bibfield{author}{\bibinfo{person}{Deze Wang}, \bibinfo{person}{Zhouyang Jia}, \bibinfo{person}{Shanshan Li}, \bibinfo{person}{Yue Yu}, \bibinfo{person}{Yun Xiong}, \bibinfo{person}{Wei Dong}, {and} \bibinfo{person}{Xiangke Liao}.} \bibinfo{year}{2022}\natexlab{a}.
\newblock \showarticletitle{Bridging pre-trained models and downstream tasks for source code understanding}. In \bibinfo{booktitle}{\emph{Proceedings of the 44th International Conference on Software Engineering}}. \bibinfo{pages}{287--298}.
\newblock


\bibitem[Wang et~al\mbox{.}(2022b)]%
        {wang2022jtrans}
\bibfield{author}{\bibinfo{person}{Hao Wang}, \bibinfo{person}{Wenjie Qu}, \bibinfo{person}{Gilad Katz}, \bibinfo{person}{Wenyu Zhu}, \bibinfo{person}{Zeyu Gao}, \bibinfo{person}{Han Qiu}, \bibinfo{person}{Jianwei Zhuge}, {and} \bibinfo{person}{Chao Zhang}.} \bibinfo{year}{2022}\natexlab{b}.
\newblock \showarticletitle{Jtrans: Jump-aware transformer for binary code similarity detection}. In \bibinfo{booktitle}{\emph{Proceedings of the 31st ACM SIGSOFT International Symposium on Software Testing and Analysis}}. \bibinfo{pages}{1--13}.
\newblock


\bibitem[Wang and Wu(2017)]%
        {IMF}
\bibfield{author}{\bibinfo{person}{Shuai Wang} {and} \bibinfo{person}{Dinghao Wu}.} \bibinfo{year}{2017}\natexlab{}.
\newblock \showarticletitle{In-Memory Fuzzing for Binary Code Similarity Analysis}. In \bibinfo{booktitle}{\emph{Proceedings of the 32nd IEEE/ACM International Conference on Automated Software Engineering}} (Urbana-Champaign, IL, USA) \emph{(\bibinfo{series}{ASE 2017})}. \bibinfo{publisher}{IEEE Press}, \bibinfo{pages}{319–330}.
\newblock
\showISBNx{9781538626849}


\bibitem[Wang et~al\mbox{.}(2021)]%
        {wang-etal-2021-codet5}
\bibfield{author}{\bibinfo{person}{Yue Wang}, \bibinfo{person}{Weishi Wang}, \bibinfo{person}{Shafiq Joty}, {and} \bibinfo{person}{Steven~C.H. Hoi}.} \bibinfo{year}{2021}\natexlab{}.
\newblock \showarticletitle{{C}ode{T}5: Identifier-aware Unified Pre-trained Encoder-Decoder Models for Code Understanding and Generation}. In \bibinfo{booktitle}{\emph{Proceedings of the 2021 Conference on Empirical Methods in Natural Language Processing}}. \bibinfo{publisher}{Association for Computational Linguistics}, \bibinfo{address}{Online and Punta Cana, Dominican Republic}, \bibinfo{pages}{8696--8708}.
\newblock
\urldef\tempurl%
\url{https://doi.org/10.18653/v1/2021.emnlp-main.685}
\showDOI{\tempurl}


\bibitem[Wu et~al\mbox{.}(2023)]%
        {wu2023effective}
\bibfield{author}{\bibinfo{person}{Yi Wu}, \bibinfo{person}{Nan Jiang}, \bibinfo{person}{Hung~Viet Pham}, \bibinfo{person}{Thibaud Lutellier}, \bibinfo{person}{Jordan Davis}, \bibinfo{person}{Lin Tan}, \bibinfo{person}{Petr Babkin}, {and} \bibinfo{person}{Sameena Shah}.} \bibinfo{year}{2023}\natexlab{}.
\newblock \showarticletitle{How Effective Are Neural Networks for Fixing Security Vulnerabilities}.
\newblock \bibinfo{journal}{\emph{arXiv preprint arXiv:2305.18607}} (\bibinfo{year}{2023}).
\newblock


\bibitem[Wu et~al\mbox{.}(2021)]%
        {wu2021representing}
\bibfield{author}{\bibinfo{person}{Zhanghao Wu}, \bibinfo{person}{Paras Jain}, \bibinfo{person}{Matthew Wright}, \bibinfo{person}{Azalia Mirhoseini}, \bibinfo{person}{Joseph~E Gonzalez}, {and} \bibinfo{person}{Ion Stoica}.} \bibinfo{year}{2021}\natexlab{}.
\newblock \showarticletitle{Representing long-range context for graph neural networks with global attention}.
\newblock \bibinfo{journal}{\emph{Advances in Neural Information Processing Systems}}  \bibinfo{volume}{34} (\bibinfo{year}{2021}), \bibinfo{pages}{13266--13279}.
\newblock


\bibitem[Xu et~al\mbox{.}(2022)]%
        {xu2022systematic}
\bibfield{author}{\bibinfo{person}{Frank~F Xu}, \bibinfo{person}{Uri Alon}, \bibinfo{person}{Graham Neubig}, {and} \bibinfo{person}{Vincent~Josua Hellendoorn}.} \bibinfo{year}{2022}\natexlab{}.
\newblock \showarticletitle{A systematic evaluation of large language models of code}. In \bibinfo{booktitle}{\emph{Proceedings of the 6th ACM SIGPLAN International Symposium on Machine Programming}}. \bibinfo{pages}{1--10}.
\newblock


\bibitem[Xu et~al\mbox{.}(2021)]%
        {xu2021mff}
\bibfield{author}{\bibinfo{person}{Guangquan Xu}, \bibinfo{person}{Meiqi Feng}, \bibinfo{person}{Litao Jiao}, \bibinfo{person}{Jian Liu}, \bibinfo{person}{Hong-Ning Dai}, \bibinfo{person}{Ding Wang}, \bibinfo{person}{Emmanouil Panaousis}, {and} \bibinfo{person}{Xi Zheng}.} \bibinfo{year}{2021}\natexlab{}.
\newblock \showarticletitle{MFF-AMD: multivariate feature fusion for Android malware detection}. In \bibinfo{booktitle}{\emph{Collaborative Computing: Networking, Applications and Worksharing: 17th EAI International Conference, CollaborateCom 2021, Virtual Event, October 16-18, 2021, Proceedings, Part I 17}}. Springer, \bibinfo{pages}{368--385}.
\newblock


\bibitem[Xu et~al\mbox{.}(2023a)]%
        {xu2023diemph}
\bibfield{author}{\bibinfo{person}{Xiangzhe Xu}, \bibinfo{person}{Shiwei Feng}, \bibinfo{person}{Yapeng Ye}, \bibinfo{person}{Guangyu Shen}, \bibinfo{person}{Zian Su}, \bibinfo{person}{Siyuan Cheng}, \bibinfo{person}{Guanhong Tao}, \bibinfo{person}{Qingkai Shi}, \bibinfo{person}{Zhuo Zhang}, {and} \bibinfo{person}{Xiangyu Zhang}.} \bibinfo{year}{2023}\natexlab{a}.
\newblock \showarticletitle{Improving Binary Code Similarity Transformer Models by Semantics-Driven Instruction Deemphasis}. In \bibinfo{booktitle}{\emph{Proceedings of the 32nd ACM SIGSOFT International Symposium on Software Testing and Analysis}} (Seattle, WA, USA) \emph{(\bibinfo{series}{ISSTA 2023})}. \bibinfo{publisher}{Association for Computing Machinery}, \bibinfo{address}{New York, NY, USA}, \bibinfo{pages}{1106–1118}.
\newblock
\showISBNx{9798400702211}
\urldef\tempurl%
\url{https://doi.org/10.1145/3597926.3598121}
\showDOI{\tempurl}


\bibitem[Xu et~al\mbox{.}(2023b)]%
        {xu2023pem}
\bibfield{author}{\bibinfo{person}{Xiangzhe Xu}, \bibinfo{person}{Zhou Xuan}, \bibinfo{person}{Shiwei Feng}, \bibinfo{person}{Siyuan Cheng}, \bibinfo{person}{Yapeng Ye}, \bibinfo{person}{Qingkai Shi}, \bibinfo{person}{Guanhong Tao}, \bibinfo{person}{Le Yu}, \bibinfo{person}{Zhuo Zhang}, {and} \bibinfo{person}{Xiangyu Zhang}.} \bibinfo{year}{2023}\natexlab{b}.
\newblock \showarticletitle{PEM: Representing Binary Program Semantics for Similarity Analysis via a Probabilistic Execution Model}.
\newblock \bibinfo{journal}{\emph{arXiv preprint arXiv:2308.15449}} (\bibinfo{year}{2023}).
\newblock


\bibitem[Yu et~al\mbox{.}(2022)]%
        {yu2022deepdi}
\bibfield{author}{\bibinfo{person}{Sheng Yu}, \bibinfo{person}{Yu Qu}, \bibinfo{person}{Xunchao Hu}, {and} \bibinfo{person}{Heng Yin}.} \bibinfo{year}{2022}\natexlab{}.
\newblock \showarticletitle{DeepDi: Learning a Relational Graph Convolutional Network Model on Instructions for Fast and Accurate Disassembly}. In \bibinfo{booktitle}{\emph{31st USENIX Security Symposium (USENIX Security 22)}}. \bibinfo{pages}{2709--2725}.
\newblock


\bibitem[Zhang et~al\mbox{.}(2019)]%
        {zhang2019bda}
\bibfield{author}{\bibinfo{person}{Zhuo Zhang}, \bibinfo{person}{Wei You}, \bibinfo{person}{Guanhong Tao}, \bibinfo{person}{Guannan Wei}, \bibinfo{person}{Yonghwi Kwon}, {and} \bibinfo{person}{Xiangyu Zhang}.} \bibinfo{year}{2019}\natexlab{}.
\newblock \showarticletitle{BDA: practical dependence analysis for binary executables by unbiased whole-program path sampling and per-path abstract interpretation}.
\newblock \bibinfo{journal}{\emph{Proceedings of the ACM on Programming Languages}} \bibinfo{volume}{3}, \bibinfo{number}{OOPSLA} (\bibinfo{year}{2019}), \bibinfo{pages}{1--31}.
\newblock


\bibitem[Zhu et~al\mbox{.}(2020)]%
        {zhu2020beyond}
\bibfield{author}{\bibinfo{person}{Jiong Zhu}, \bibinfo{person}{Yujun Yan}, \bibinfo{person}{Lingxiao Zhao}, \bibinfo{person}{Mark Heimann}, \bibinfo{person}{Leman Akoglu}, {and} \bibinfo{person}{Danai Koutra}.} \bibinfo{year}{2020}\natexlab{}.
\newblock \showarticletitle{Beyond homophily in graph neural networks: Current limitations and effective designs}.
\newblock \bibinfo{journal}{\emph{Advances in neural information processing systems}}  \bibinfo{volume}{33} (\bibinfo{year}{2020}), \bibinfo{pages}{7793--7804}.
\newblock


\bibitem[Zhu et~al\mbox{.}(2023)]%
        {zhu2023ktrans}
\bibfield{author}{\bibinfo{person}{Wenyu Zhu}, \bibinfo{person}{Hao Wang}, \bibinfo{person}{Yuchen Zhou}, \bibinfo{person}{Jiaming Wang}, \bibinfo{person}{Zihan Sha}, \bibinfo{person}{Zeyu Gao}, {and} \bibinfo{person}{Chao Zhang}.} \bibinfo{year}{2023}\natexlab{}.
\newblock \showarticletitle{kTrans: Knowledge-Aware Transformer for Binary Code Embedding}.
\newblock \bibinfo{journal}{\emph{arXiv preprint arXiv:2308.12659}} (\bibinfo{year}{2023}).
\newblock


\end{thebibliography}

\clearpage
\newpage

\appendix
\section{Implementation and Pre-training}
\label{appendix:pretrain}

\toolname has 12 transformer blocks with 768 hidden dimensions and 12 heads for each multi-head attention module, which is the same as BERT. For the relative positional bias, we set  $r_{\text{max}}$ to 8. We pre-train using the same BinaryCorps dataset as in JTrans~\cite{wang2022jtrans}, with 26 million binary functions. We set the max pre-training context length to be 512 tokens. We use the AdamW~\cite{loshchilov2017decoupled} optimizer to update model parameters with a batch size of 1024 and a 5e-4 learning rate. We set the warm-up steps to be 10k and use a linear learning rate scheduler to schedule the learning rate. The model is pre-trained from scratch without initialization from any pre-trained checkpoints. The pre-training takes 25k steps to converge.

We pre-train the model on a Linux cluster with 8 Nvidia Tesla A100 GPUs each with 80GB memory. Our implementation of the model is based on Huggingface transformers~\footnote{https://github.com/huggingface/transformers} version 4.30.2 and PyTorch 2.0.1 stable~\footnote{https://pytorch.org/}.

\section{Malware Family Classification Details}
\label{appendix:mfc}

\begin{figure}
    \centering
    \includegraphics[width=0.8\linewidth]{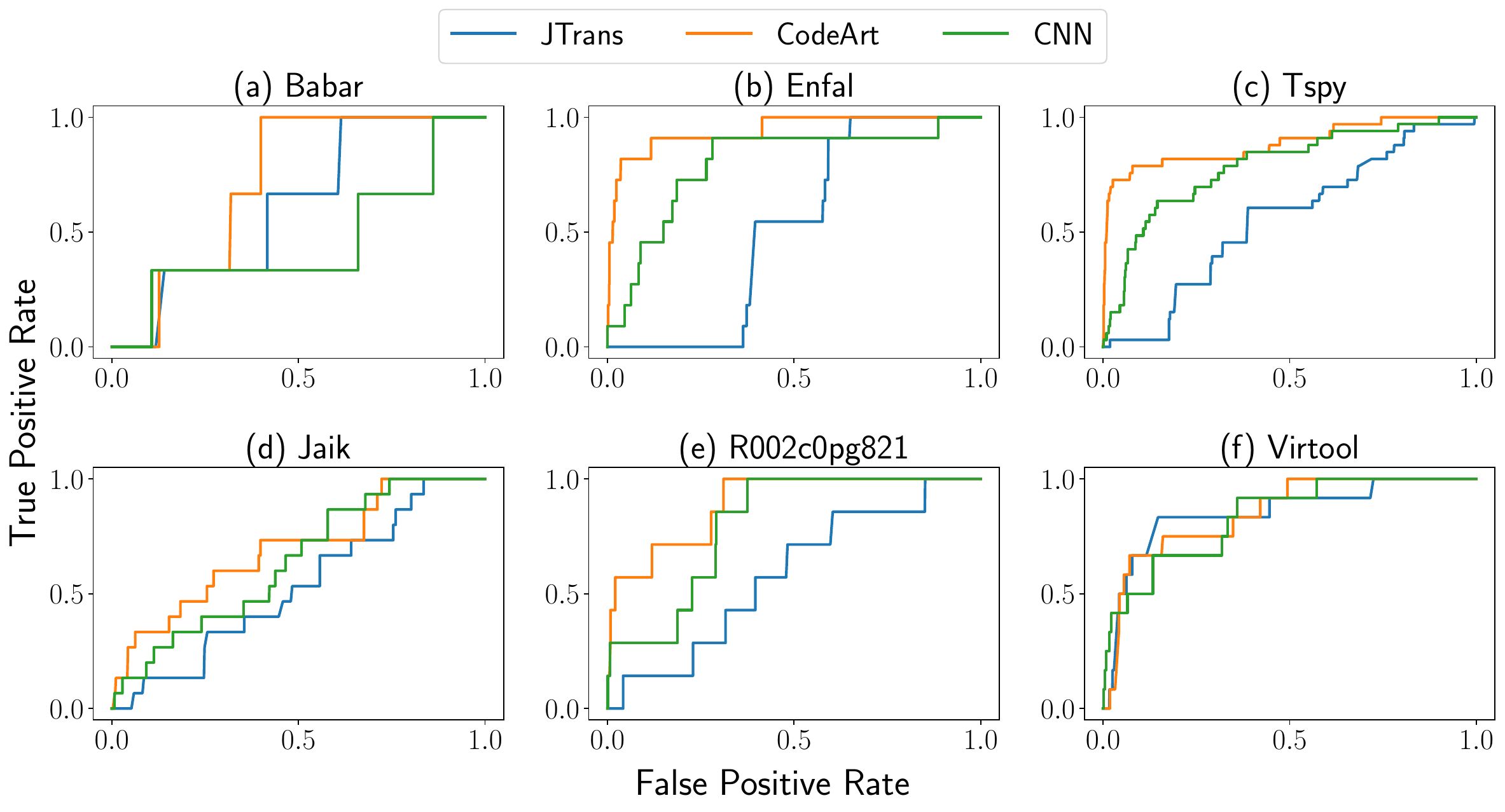}
    \caption{ROC Curves for CNN, \toolname, and JTrans models encoding the forefront 2 functions.}
    \label{fig:mal-roc}
\end{figure}

\smallskip
\noindent
{\bf Metric Details.}
Consider $n_s$ samples and $n_l$ labels for a multi-label evaluation. Given a matrix of ground-truth labels $y\in\{0,1\}^{n_s\times n_l}$ and the predicted score associated with each label $\hat{f}\in\mathbb{R}^{n_s\times n_l}$, the LRAP and LRL~\cite{tsoumakas2010mining} are computed as follows. 

\begin{equation}
    \text{LRAP}(y, \hat{f})=\frac{1}{n_s}\sum_{i=0}^{n_s-1}\frac{1}{\|y_i\|_0}\sum_{j:y_{ij}=1}\frac{|\mathcal{L}_{ij}|}{\text{rank}_{ij}}
\end{equation}

\begin{equation}
    \text{LRL}(y,\hat{f}) = \frac{1}{n_s}\sum_{i=0}^{n_s-1}\frac{1}{\|y_i\|_0(n_l - \|y_i\|_0)}\big|\{(k,l):\hat{f}_{ik}\leq\hat{f}_{il},y_{ik}=1,y_{il}=0\}\big|
\end{equation}

where $\mathcal{L}_{ij}=\{k:y_{ik}=1,\hat{f}_{ik}\geq \hat{f}_{ij}\}$, $\text{rank}_{ij}=\big|\{k:\hat{f}_{ik}\geq\hat{f}_{ij}\}\big|$, $|\cdot|$ is the set cardinality, and $\|\cdot\|_0$ is the $l_0$-normalization.

\smallskip
\noindent
{\bf Class-wise Results.}
We further show the per-class ROC curves for 6 randomly sampled malware families in Fig.~\ref{fig:mal-roc}. The area under \toolname's ROC curve is significantly greater than that of JTrans and CNN in \emph{Babar}, \emph{Enfal}, \emph{Tspy}, \emph{R002c0pg821}, and \emph{Jaik}. For \emph{Virtool}, \toolname's performance is similar to the other two. This demonstrates that \toolname excels in classification for each family.

\section{GraphCodeBERT-like Pre-training}
\label{appendix:gcb-like}

\begin{figure}
    \centering
    \includegraphics[width=0.9\linewidth]{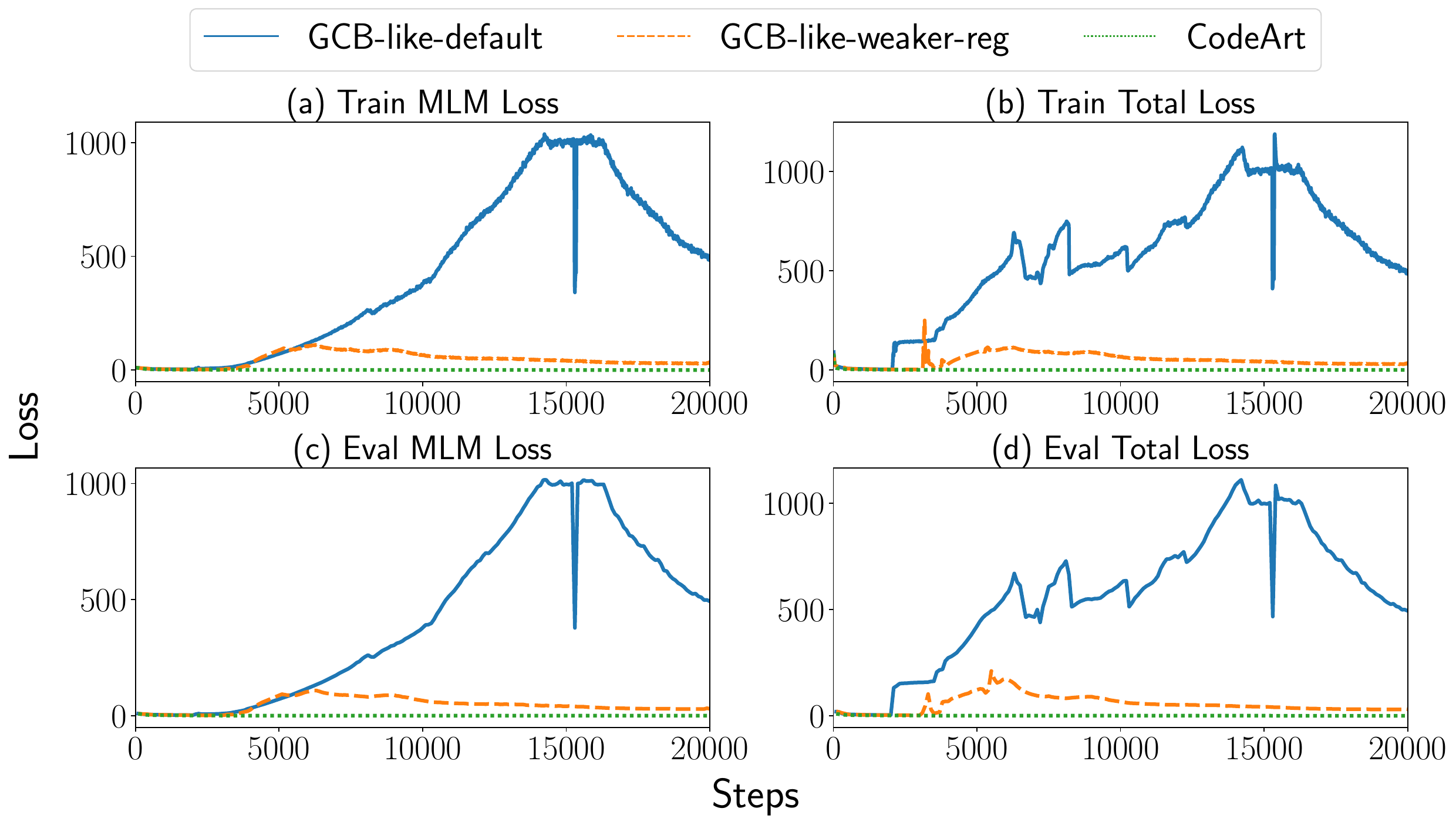}
    \caption{Training curves for GraphCodeBERT-like pre-training. The x-axes denote training steps, and the y-axes denote the values of loss. {\em GCB-like-default} denotes the model trained with the default GCB loss terms. {\em GCB-like-weaker-reg} denotes the model trained with a smaller weight for
    data-flow loss.
    {\em MLM Loss} denotes the loss for masked language modeling and {\em Total Loss} the sum of the MLM loss and the data-flow loss.
    }
    \label{fig:gcb-curve}
\end{figure}

\smallskip\noindent
{\bf GCB-like Pre-training.}
GraphCodeBERT~\cite{guo2020graphcodebert} on source code augmented the input source code sequence with additional data-flow related variable tokens and used them as graph nodes for graph modeling. We mimic this for assembly code by augmenting the input instruction sequence with additional operand tokens that correspond to operands in individual instructions. These operand tokens serve as dependence graph nodes for the GCB-like pre-training, which can be denoted as $V=\{v_1,\cdots, v_k\}$. We denote $E=\{e_1,\cdots, e_l\}$ as the \emph{directed} dependence edges for these nodes.

Then, we mimic the original GraphCodeBERT to introduce the graph related objectives with these nodes and edges. For edge prediction, we randomly sample 20\% of nodes $V_s\in V$ and then sample a balanced number of edges from $(V_s\times V \cup V\times V_s) \cap E$ and $V_s\times V \cup V\times V_s - E$, resulting in $E_c$. Then, the edge prediction loss can be computed as follows.

\begin{equation}
    \mathcal{L}_{\text{EdgePred}} = -\sum_{e_{ij}\in E_c}[{\mathbb{I}(e_{ij}\in E})\log p_{e_{ij}} + (1-\mathbb{I}(e_{ij}\in E))\log (1-p_{e_{ij}})],
\end{equation}

where $e_{ij}$ is the dot product of the embeddings of node $i$ and node $j$ converted into a probability by the sigmoid function.

Node alignment is essentially another edge prediction between node tokens and the corresponding operand tokens in the original input sequence. Let $C=\{c_1,\cdots, c_m\}$ be the input token sequence and $E_{\text{align}}$  the set of edges between nodes and the corresponding operand tokens. 20\% of randomly sampled nodes form $V_s'$. Edges are sampled in a balanced way from $E_{\text{align}}$ and $V'_s\times C - E_{\text{align}}$, resulting in $E_c'$. The node alignment loss can be computed as follows.

\begin{equation}
    \mathcal{L}_{\text{NodeAlign}} = 
    -\sum_{e_{ij}\in E_{c}'}[\mathbb{I}(e_{ij}\in E_{\text{align}})\log p_{e_{ij}} 
    + (1-\mathbb{I}(e_{ij}\in E_{\text{align}}))\log (1-p_{e_{ij}})]
\end{equation}

where $e_{ij}$ is computed similarly as for edge prediction loss.

We also implement the graph-guided mask for GCB-like following the original paper, which works like regular mask for multi-head attention. Note that the graph-guided mask is completely different from our attention mask in that it only prohibits (1) signals from node tokens to input tokens except for the corresponding ones, and (2) signals from non-dependent node tokens. Therefore, for the graph-guided mask, the original input tokens can freely attend to each other throughout all layers. 

The MLM objective of GraphCodeBERT is exactly the same as BERT's and hence ours. The final objective is the sum of MLM objective and the weighted additional objectives.

\begin{equation}
    \mathcal{L}_{\text{GCB-like-default}} = \mathcal{L}_{\text{MLM}} + \lambda(\mathcal{L}_{\text{EdgePred}} + \mathcal{L}_{\text{NodeAlign}})
\end{equation}

For GCB-like-default, we set $\lambda=1$, and for GCB-like-weaker-reg, we set $\lambda=0.1$.

The architecture for GCB-like is just the default BERT and the rest of pre-training setting is the same as \toolname.

\noindent
{\bf Optimization Stability.} We show the training curves of both \toolname and GraphCodeBERT-like pre-training in Fig.~\ref{fig:gcb-curve}. For GCB-like-default, we can see that the additional loss introduced by GCB caused instability (increasing loss) when optimizing the MLM loss and therefore the total loss after 5000 steps. We speculate it is because binary code has significantly fewer distinct variable names 
than source code. Without attention regularization,
the additional losses in GCB 
eventually conflict with the MLM loss.
For {\em GCB-like-weaker-reg}, if we reduce the weights for the data-flow
loss, the training can converge but it still suffers from instability. 
In comparison, the training curve of \toolname is significantly more stable than both variants of GCB. It demonstrates the effectiveness of our attention regularization.

\end{document}